%% file: STO_Oct23.tex
\documentclass[prb,prb,reprint,aps,showpacs,showkeys,twocolumn]{revtex4-1}

\pdfoutput=1

\usepackage[utf8]{inputenc}
\usepackage[T1]{fontenc}
\usepackage{hyperref}
\usepackage{color}
\usepackage{graphicx}
\usepackage{units}
\usepackage{amsmath}

\hypersetup{colorlinks = true, citecolor = blue, breaklinks = true}

\begin{document}

\title{Competition and cooperation between antiferrodistortive and ferroelectric instabilities in SrTiO$_3$}
\date{\today}
\author{Ulrich Aschauer}
\affiliation{Materials Theory, ETH Zurich, Wolfgang-Pauli-Strasse 27, CH-8093 Z\"urich, Switzerland}
\author{Nicola A. Spaldin}
\affiliation{Materials Theory, ETH Zurich, Wolfgang-Pauli-Strasse 27, CH-8093 Z\"urich, Switzerland}

\begin{abstract}
We use density functional theory to explore the interplay between octahedral rotations and ferroelectricity in the model compound SrTiO$_3$. We find that over the experimentally relevant range octahedral rotations suppress ferroelectricity as is generally assumed in the literature. Somewhat surprisingly we observe that at larger angles the previously weakened ferroelectric instability strengthens significantly. By analysing geometry changes, energetics, force constants and charges, we explain the mechanisms behind this transition from competition to cooperation with increasing octahedral rotation angle.
\end{abstract}

\maketitle

\section{Introduction}

Materials with the ABO$_3$ perovskite crystal structure display a multitude of interesting properties of fundamental and technological importance such as ferroelectricity \cite{Rabe:2007up}, superconductivity \cite{Bednorz:1988bw} or metal-insulator transitions \cite{Ramirez:1997uh} and have been at the focus of intense research for many years. Most perovskites structurally distort from the perfect cubic $Pm\overline{3}m$ structure and their properties are strongly affected or even determined by these distortions. Two of the possible distortions are antiferrodistortive (AFD: rotations of the oxygen octahedra) and ferroelectric (FE: polar displacement of cation and anion sublattices against one another).

The observation that the large majority of perovskites contain AFD octahedral rotations \cite{Woodward:1997ci,Woodward:1997js} but that these distortions are absent in the most common ferroelectric perovskites such as BaTiO$_3$ and PbTiO$_3$ has led to the general opinion that octahedral rotations suppress ferroelectricity in perovskites. Theoretical studies on selected materials further support this point, as reducing the magnitude of octahedral rotations via tensile strain or artificially disabling them has been shown to induce ferroelectricity in several cases \cite{Eklund:2009hp,Bhattacharjee:2009eb,Hatt:2010fi,Rondinelli:2009fk}. One example recently studied in detail is the class of perovskites with the \textit{Pnma} space group \cite{Benedek:2013jl}, in which it was shown that octahedral rotations cause anti-polar displacements of the A-site, which in turn suppress the ferroelectric instability. It is however still unclear if  AFD and FE instabilities are genuinely competing in all substructure types and if so, what the underlying microscopic mechanism would be.

\begin{figure}
\centering
\def\svgwidth{\columnwidth}
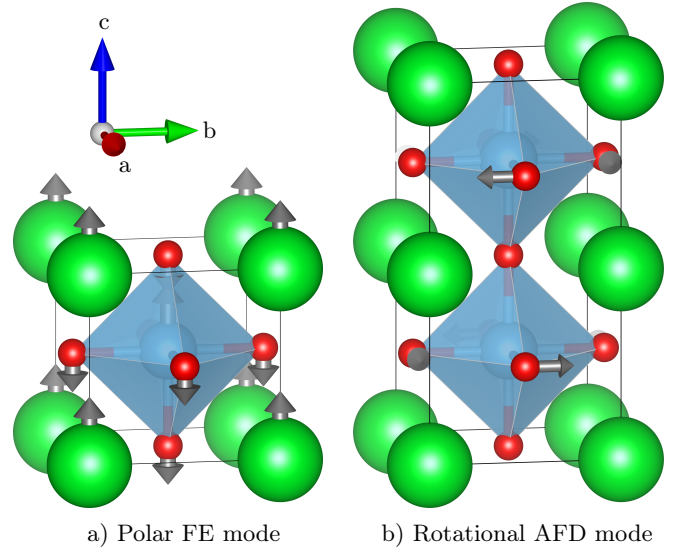
\caption{\label{fig:structure} Symmetry lowering distortions in SrTiO$_3$ considered in this work: a) polar (FE) displacements along the c axis and b) antiferrodistortive (AFD) rotations around the c axis. Directions of atomic displacements in the cubic phase are indicated by gray arrows of arbitrary length. Actual displacement magnitudes are shown in Fig. \ref{fig:displacements}a). Color code: Sr = green, Ti = blue, O = red.}
\end{figure}

In the present work we investigate the existence of this competition and its microscopic mechanism in the model compound SrTiO$_3$ (STO). STO exhibits antiferrodistortive octahedral rotations below 105K \cite{Shirane:1969ug,Cowley:1969ul} described as a$^0$a$^0$c$^-$ in the Glazer notation, \cite{Glazer:1972eb} i.e. the BO$_6$ octahedra rotate around the c axis with rotations in consecutive layers being out of phase (Fig. \ref{fig:structure}b). STO is also known to be an incipient (quantum) paraelectric \cite{Zhong:1996vq,Vanderbilt:1998fs}, meaning that it shows a ferroelectric instability (Fig. \ref{fig:structure}a), which is too weak to manifest itself even at the lowest attainable temperatures \cite{Muller:1979wa} due to quantum fluctuations. The relative structural simplicity and the co-existence of both instabilities make STO a good prototype material to study the competition between the AFD and the FE instability. In addition, since it is not a member of the \textit{Pnma} (a$^-$a$^-$c$^+$) class, it allows us to study if the competition between FE and AFD exists also in the absence of anti-polar A site displacements  \cite{Benedek:2013jl}.

\section{Computational details}

We perform density functional theory (DFT) calculations using the Vienna \textit{ab initio} simulation package (VASP) \cite{Kresse:1993ty,Kresse:1994us,Kresse:1996vk,Kresse:1996vf} with the supplied PAW potentials \cite{Blochl:1994uk,Kresse:1999wc} considering Sr(4s, 4p, 5s), Ti(3p, 3d, 4s) and O(2s, 2p) valence electrons. Calculations were performed within the local density approximation (LDA) as well as using the gradient corrected PBE \cite{Perdew:1996iq} and PBEsol \cite{Perdew:2008fa} density functionals and the performance of the individual functionals will be discussed below. Electronic wavefunctions were expanded in planewaves up to a kinetic energy cutoff of 550 eV and the reciprocal space was sampled using a shifted $8 \times 8 \times 8$ Monkhorst-Pack \cite{Monkhorst:1976ta} mesh for the five-atom unit-cell and the corresponding $4 \times 4 \times 4$ mesh for the $2 \times 2 \times 2$ supercells. Structures and cells were relaxed until forces converged below 10$^{-5}$ eV/\AA. AFD rotations were imposed during structural optimisation by freezing the in-plane component of the position of the equatorial oxygen atoms. Phonon calculations were then performed using the frozen-phonon method within the phonopy code \cite{Togo:2008jt} and force constant matrices rotated for analysis so that the $xx$ components lay along the appropriate bond axis.

\section{Results \& Discussion}

\subsection{Effect of the density functional}

\begin{figure}
\centering
\def\svgwidth{\columnwidth}
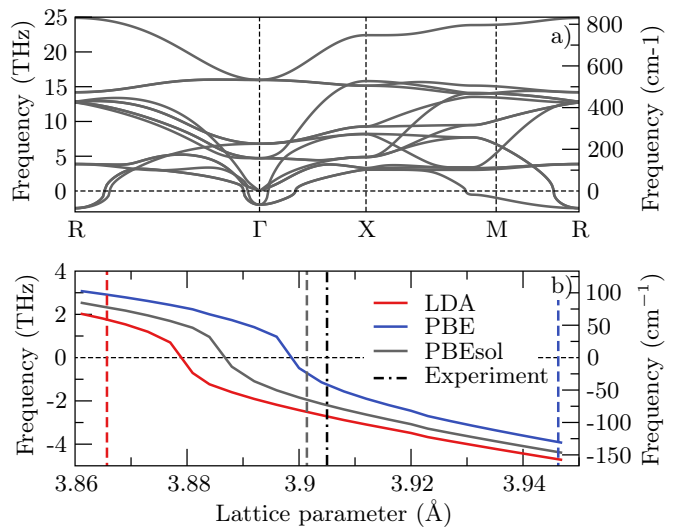
\caption{\label{fig:instabilities}a) Phonon band-structure computed using the PBEsol functional and b) FE soft mode frequency as a function of the lattice parameter for different functionals. In these plots negative frequencies are used to represent imaginary frequencies.} 
\end{figure}

Fig. \ref{fig:instabilities}a) shows the phonon dispersion for the high symmetry cubic phase computed using the PBEsol exchange-correlation functional, imaginary phonon frequencies being plotted as negative values. We see imaginary modes at the R point of the Brillouin zone, corresponding to the antiferrodistortive (AFD) octahedral rotation described by the a$^0$a$^0$c$^-$ Glazer tilt pattern as shown in Fig. \ref{fig:structure}b). The unstable mode at the M point corresponds to the energetically less favourable a$^0$a$^0$c$^+$ Glazer tilt, which we will not consider any further in this work. The unstable modes at the $\Gamma$ point are the ferroelectric (FE) polar modes shown in Fig. \ref{fig:structure}a).

While the R and M point instabilities are fairly insensitive to the choice of functional and volume, the polar instability exhibits a strong volume and functional dependence, which is shown in Fig. \ref{fig:instabilities}b). At small volumes no FE instability exists for any of the functionals, whereas upon increasing the lattice parameter by merely 0.08 \AA, all functionals predict a FE instability. It is interesting to relate these results to the calculated equilibrium lattice parameters shown as vertical dashed lines. LDA at the LDA equilibrium volume predicts STO not to have a FE instability. The gradient corrected functionals predict the existence of this instability at their respective equilibrium volume, PBE having a stronger instability than PBEsol. Comparing to the experimental lattice parameter, LDA and PBE respectively show the usual under- and overestimation of the lattice parameter. PBEsol on the other hand results in fairly good agreement with the experimental lattice parameter. Given that in the present work we wish to relax the cell dimensions so as to obtain the tetragonal cell shape resulting from both AFD and FE distortions, the PBEsol functional represents the best choice. Therefore all calculations described hereafter were carried out using the PBEsol functional. We note however that all semi-local functionals (without or with +U correction) overestimate the equilibrium AFD angle and to obtain good agreement with experiment for this quantity hybrid functionals are a better choice \cite{Wahl:2008ea, Evarestov:2011iv, ElMellouhi:2013bu}. We will briefly return to the reason behind this point in section \ref{section:sojt}.

\subsection{Existence of the competition and cooperation}

\begin{figure}
\centering
\def\svgwidth{\columnwidth}
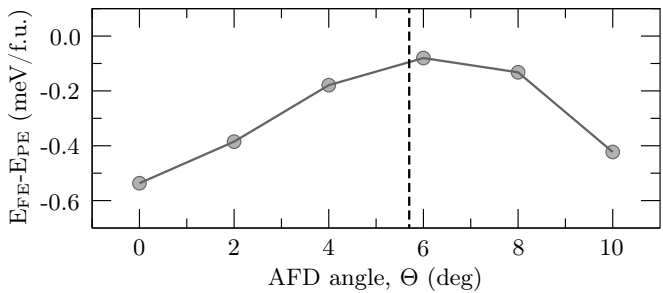
\caption{\label{fig:competition}Energy gain due to the FE distortion as a function of the AFD angle. The dashed vertical line indicates the theoretical (PBEsol) equilibrium AFD rotation angle.}
\end{figure}

In Fig. \ref{fig:competition} we show the depth of the FE energy double-well (i.e. the energy difference E$_\mathrm{FE}$ - E$_\mathrm{PE}$ between the ferroelectric (FE) and paraelectric (PE) state) as a function of the AFD angle. The FE state was reached by displacing all atoms by a small distance ($\sim$ 0.02 \AA) along the bulk FE eigenvector and relaxing the cell and internal coordinates while keeping the fractional in-plane coordinates of the equatorial O fixed to impose the AFD angle. In the cubic structure (at 0$^\circ$ ADF angle) we calculate that STO is FE with an energy gain of 0.8 meV per formula unit. As the AFD amplitude increases, the energy gain associated with the FE distortion becomes smaller, reaching around 0.2 meV per formula unit at the theoretical equilibrium AFD angle of 5.7$^\circ$. This small energy gain is consistent with the material's observed quantum paraelectricity, quantum fluctuations being sufficient to suppress the FE instability. The energy gain for the FE distortion then increases again rapidly and at 10$^\circ$ is already almost as large as it was in the cubic structure. We note here that the FE mode frequency (not shown) follows the same trend as the well depth. The AFD thus renders the FE instability energetically and dynamically less favourable. We see that in the experimentally relevant range of angles $\left(\lesssim6^\circ\right)$ a competition exists, as the FE instability is weakened by the AFD rotations with maximum effect around the theoretical equilibrium angle. At larger angles, the AFD and FE cooperate, the FE instability increasing beyond that found in the cubic structure. In the remainder of this article we explain the microscopic mechanisms underlying these two observations.

\subsection{Effect of geometry changes}

\begin{figure}
\centering
\def\svgwidth{\columnwidth}
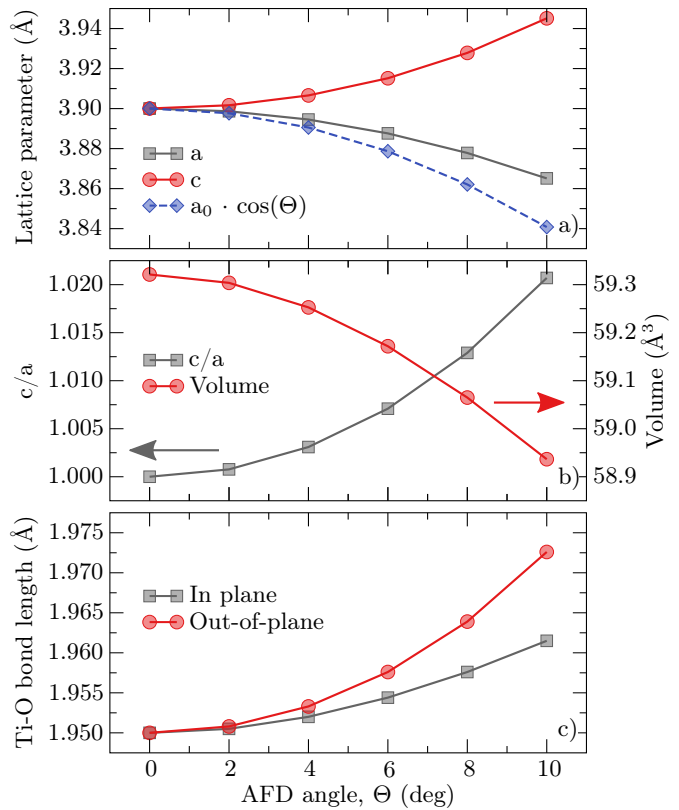
\caption{\label{fig:cell_change}Evolution of various structural parameters as a function of the AFD rotation angle without FE displacements: a) In-plane and out-of-plane lattice parameters. The dashed line indicates the evolution of the a lattice parameter for perfectly rigid TiO$_6$ octahedra. b) c/a ratio and volume. c) Ti-O bond lengths for the in-plane and out-of-plane bonds.} 
\end{figure}

Since we found the FE instability to be strongly volume dependent we next investigate whether the observed competition is solely a volume effect. We first calculate changes in lattice parameters and bond lengths induced by the AFD rotation. In Fig \ref{fig:cell_change}a) we see that, as expected for the AFD rotation, the cell becomes tetragonal, as the a=b lattice parameters decrease while the lattice expands along the c direction with increasing AFD angle. This is also reflected in the c/a ratio shown in Fig. \ref{fig:cell_change}b), which increases with increasing AFD angle. It is however interesting to note that the a=b lattice parameter in Fig. \ref{fig:cell_change}a) does not follow the dashed curve predicted for perfectly rigid TiO$_6$ octahedra, but remains larger at all angles. This implies that the Ti-O bonds of the octahedra are not rigid but stretch as the AFD angle increases. We see this to be the case in Fig. \ref{fig:cell_change}c), where both the in-plane and out-of-plane bond lengths  are shown to increase with increasing AFD angle. We also observe that the volume decreases with increasing AFD angle (Fig. \ref{fig:cell_change}b), indicating a deviation of Poisson's ratio from 0.5.

\begin{figure}
\centering
\def\svgwidth{\columnwidth}
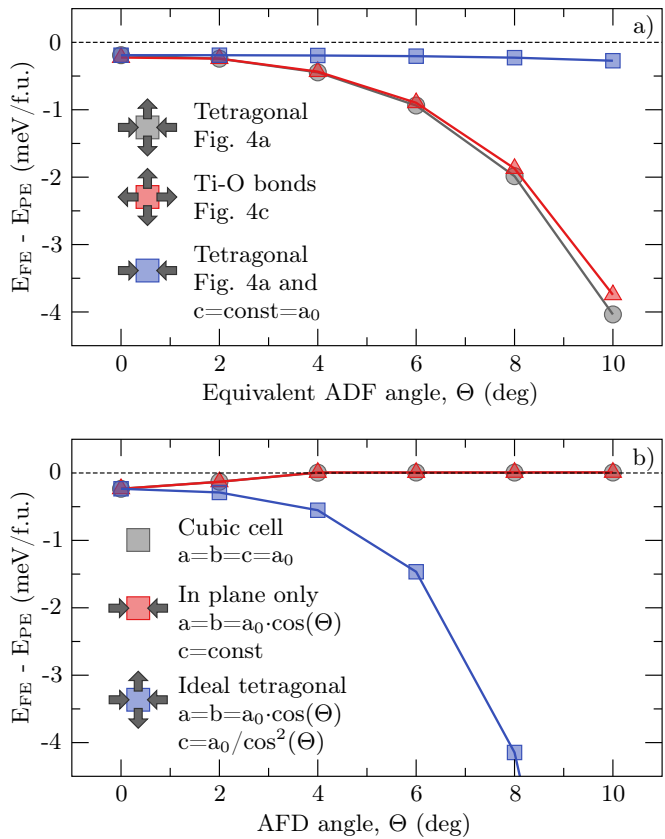
\caption{\label{fig:experiment}Computer experiments for a) cell shape changes in absence of AFD rotations and b) cell shape changes in presence of AFD rotations. In a) ``Equivalent AFD angle'' refers to the AFD angle corresponding to the imposed cell parameters. See text for details.} 
\end{figure}

In order to separate the effects of the different geometry changes, we perform computer experiments to isolate them, initially in the absence of AFD rotations. To test the effect of the cell tetragonality, we impose the cell parameters of the fully relaxed cell (i.e. the lattice parameters shown in Fig. \ref{fig:cell_change}a)). To test the effect of Ti-O bond stretching, we impose cell parameters resulting in Ti-O bond lengths equal to those in the fully relaxed cell (i.e. cell parameters equal to double the Ti-O bond-lengths shown in Fig. \ref{fig:cell_change}c)). The resulting depths of the FE energy wells for these two situations are shown in Fig. \ref{fig:experiment}a). We note that, since here we do not allow the cell to increase its tetragonality as a result of the FE distortion, the energy gain at 0$^\circ$ AFD angle is smaller than in Fig. \ref{fig:competition}. For both of these computer experiments (grey and red curves in Fig. \ref{fig:experiment}) we observe an enhancement of the FE instability and these effects can hence not be at the origin of the observed reduction of FE by AFD in STO. It is interesting to note that the cell-shape mimicking distortion is associated with a decrease in volume, while the bond-length mimicking distortion is accompanied by an increase in volume. The FE instability hence seems to be fairly independent of the \textit{total} volume within the relatively small range considered here, in contrast to the larger range shown in Fig. \ref{fig:instabilities}. Given that the two curves in Fig. \ref{fig:experiment}a) are very similar, we identify a common structural parameter, which is the expanding c lattice parameter. In fact when we keep the c lattice parameter fixed at the cubic value and perform the same computation, we obtain the nearly flat blue curve in Fig. \ref{fig:experiment}a). These computer experiments in absence of AFD rotations indicate that a tetragonal cell strongly enhances FE along the expanded axis while the compression of the in-plane axes has only a very minor effect.

Thus far our computer experiments on the lattice parameters and cell shape but without AFD rotations have not found a reduction of the FE. Therefore, to gain further insight, we perform an additional series of three computer experiments, this time including AFD rotations. For the first experiment we keep the lattice fixed at the ideal cubic lattice parameters, while for the second we adjust the in-plane lattice parameters so as to keep Ti-O bond lengths constant while keeping the c axis fixed at the cubic value. Finally we perform a third series of calculations, where the in-plane lattice parameters are adjusted to keep Ti-O bonds constant and the c-axis is chosen so as to keep the volume constant. Our results are shown in Fig. \ref{fig:experiment}b). First we observe a complete suppression of FE for the perfectly cubic situation (grey curve, partly masked by the red curve). Changing the in-plane lattice parameters (red curve) results in no change of the FE well depth, which is consistent with the weak effect of the in-plane lattice parameters shown above without AFD rotations (blue curve in Fig. \ref{fig:experiment}a). If we however allow the c-axis to expand to keep the volume constant, we recover the marked enhancement of the FE with increasing AFD angle.

From the above results we conclude that expansion of the c lattice parameter, which occurs in all situations except for the perfectly cubic lattice, enhances FE \cite{Haeni:2004gj,Choi:2004it}. If this c-axis expansion is absent but AFD rotations are present, we observe a complete suppression of FE. Therefore at this point we can make the hypothesis that in the fully relaxed situation shown in Fig. \ref{fig:competition} two competing phenomena occur: firstly, the cell becomes increasingly tetragonal, which enhances FE. Within the usual second-order Jahn-Teller picture, this enhancement of the FE with increasing tetragonality can be explained by the  loss of Ti-O $\pi$-bonding along the c-axis, part of which can be recovered by the FE distortion, combined with the reduction in coulomb repulsion associated with Ti off-centering when the atoms are further apart. Secondly, some aspect related to the AFD rotations - such as changes in bonding - suppresses the FE. The former phenomenon dominates at larger angles, leading to the observed reappearance of FE, while the latter dominates at small rotation angles. We will now focus on establishing the microscopic mechanism behind the second phenomenon.

\subsection{Microscopic mechanism}

\begin{figure}
\centering
\def\svgwidth{\columnwidth}
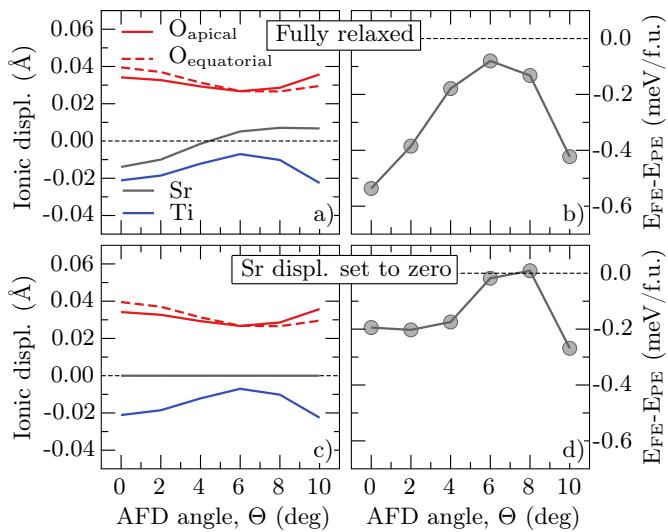
\caption{\label{fig:displacements}a) Ionic displacement associated with the FE distortion as a function of the AFD angle and b) the resulting evolution of the FE double-well depth. In c) we have reset the Sr ions to their high-symmetry position and again show the ionic displacements and d) the resulting depth of the FE double-well.}
\end{figure}

In non-lone-pair active perovskites, it is generally assumed that the FE instability is driven by an optimisation of the B-site coordination (in terms of both geometry and bonding), whereas octahedral rotations are driven by an optimisation of the A-site coordination \cite{Woodward:1997js}. We see however from the displacement pattern shown in Fig. \ref{fig:displacements} that in the structure without AFD rotations the displacements of Sr in the FE distortion are significant and comparable to those of Ti, which is also reflected in the anomalously high Sr Born effective charges computed for STO \cite{Zhong:1994ui,Lasota:1997dq}. This indicates that in the cubic structure the FE distortion simultaneously optimises the A and B-site coordination. With increasing AFD angle the Sr displacement decreases and even changes sign, being in the same direction as the oxygen displacements at AFD angles larger than $\sim$5$^\circ$. At the same time, the displacement of the O atoms also decreases, more markedly for the equatorial than for the apical atoms. The AFD significantly shortens some of the distances between equatorial O and Sr, optimising the A-site coordination. As the A-site has already optimised its coordination environment due to the rotation, the driving force to further do so via the FE distortion is reduced. This is manifested in the decrease of all FE displacement amplitudes with increasing AFD angle up to about 6$^\circ$. At larger angles the FE distortion becomes energetically increasingly favourable and the Ti and O displacement amplitudes increase again. This is in agreement with the increased tetragonality of the cell, which is analogous to STO under compressive epitaxial strain \cite{Haeni:2004gj}.

\begin{figure}
\centering
\def\svgwidth{\columnwidth}
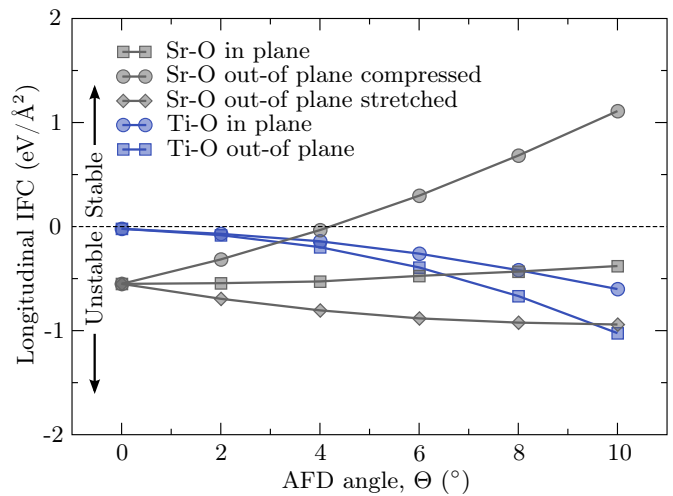
\caption{\label{fig:ifc}Longitudinal interatomic force constants (IFC) for the most strongly AFD-dependent nearest-neighbor bonds. Other interactions, such as next-nearest neighbour Ti-O have been omitted even though more unstable than the ones shown here, as they do not change with the AFD.}
\end{figure}

The evolution of the cation displacement patterns is in agreement with the longitudinal interatomic force constants (IFC), shown in Fig. \ref{fig:ifc}. Negative values indicate unstable IFCs (the two sublattices displace against one another) and positive values stable IFCs (the two sublattices displace in the same direction). The IFCs show that the nearest neighbour Ti-O instability is fairly weak in the cubic phase and that the Sr-O instability is initially much stronger. As the AFD angle increases, part of the Sr-O instability is lost while the Ti-O instability increases. The FE instability thus has an important Sr contribution in the cubic phase but becomes increasingly Ti dominated at larger AFD angles. The compressed out-of plane Sr-O IFCs become markedly less unstable with increasing AFD angle, changing from unstable to stable at around 4$^\circ$ AFD angle. This angle roughly coincides with that where Sr starts moving in the same direction as O in Fig. \ref{fig:displacements}a), which is consistent with the crossover to stability. 

It is further interesting to note here that when IFCs are computed using LDA and PBE at the respective equilibrium volumes, the Sr-O IFCs do not vary much from functional to functional. The Ti-O IFCs however are stable in the cubic structure using LDA (+0.8 eV/\AA$^2$) but unstable with PBE (-0.95 eV/\AA$^2$). Their change as a function of AFD angle is very similar from functional to functional, meaning that LDA Ti-O IFCs become unstable only at angles larger than 8$^\circ$ and 10$^\circ$ respectively for the out-of-plane and in-plane contributions. When using PBE, Ti-O IFCs are unstable for any AFD angle and become increasingly more unstable with increasing AFD angle. This shows that the large differences between functionals stem from their description of the Ti-O interactions.

In order to further understand the importance of the Sr displacement, we have, in analogy to what was done in Ref. \cite{Benedek:2013jl}, computed the energy of the relaxed structure but with Sr ions returned to their position in the high-symmetry cubic phase. We see in Fig. \ref{fig:displacements}d) that this has a marked effect on the energy well. Its depth is reduced by almost a factor three and we now observe complete suppression of the FE instability at around 8$^\circ$ AFD angle. This shows that in the cubic structure and at small AFD angles the Sr displacements of the FE distortion lead to strong bonding optimisation and hence have an important contribution to the FE energy lowering. The situation here seems to be different from \textit{Pnma} perovskites, \cite{Benedek:2013jl} because in STO A-site displacements do not suppress the FE instability but rather  strengthen it due to their involvement in the FE distortion. This could however be related to the fact that here the A-site displacements occur along the AFD rotation axis whereas in \textit{Pnma} perovskites displacements perpendicular to the rotation axis are favoured by the geometry.

\begin{figure}
\centering
\def\svgwidth{\columnwidth}
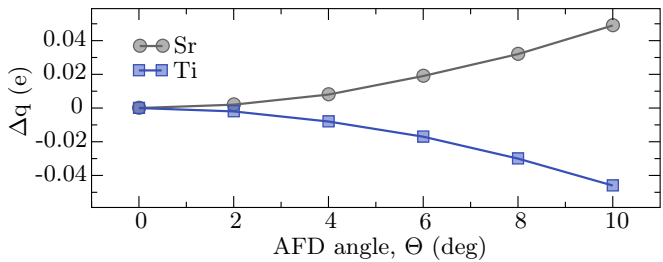
\caption{\label{fig:charges}Changes in charges obtained by integration over the PAW spheres of the cations as a function of the AFD angle.}
\end{figure}

We see in Fig. \ref{fig:displacements}a) that the Ti displacement initially decreases in magnitude with increasing AFD angle, following a similar pattern to the energy well shown in Fig. \ref{fig:displacements}b). We can understand this effect by monitoring the charges integrated over the PAW sphere of each ion. The changes in cation charges are about a factor five larger than those of the oxygen ions, therefore in Fig. \ref{fig:charges} we show only the cation charges. We see that with increasing AFD angle Ti ions lose electrons while Sr ions gain electrons. An orbital-resolved charge (not shown) reveals that the electrons flowing away from Ti originate from the most energetic hybridised orbitals, which are formed by the shorter in-plane $\sigma^*$ $\mathrm{Ti_{3d, x^2-y^2}-O_{2p}}$ orbitals. The electron flow, which is associated with enhanced Sr-O bonding thus strengthens the Ti-O bonds by weakening their anti-bonding character \cite{Woodward:1997js}. These stronger in-plane bonds oppose the Ti off-centering and hence weaken the FE instability in STO with increasing AFD angle. At larger angles the effect of increased tetragonality enhances the FE again.

\subsection{Second order Jahn-Teller viewpoint}\label{section:sojt}

\begin{figure}
\centering
\def\svgwidth{\columnwidth}
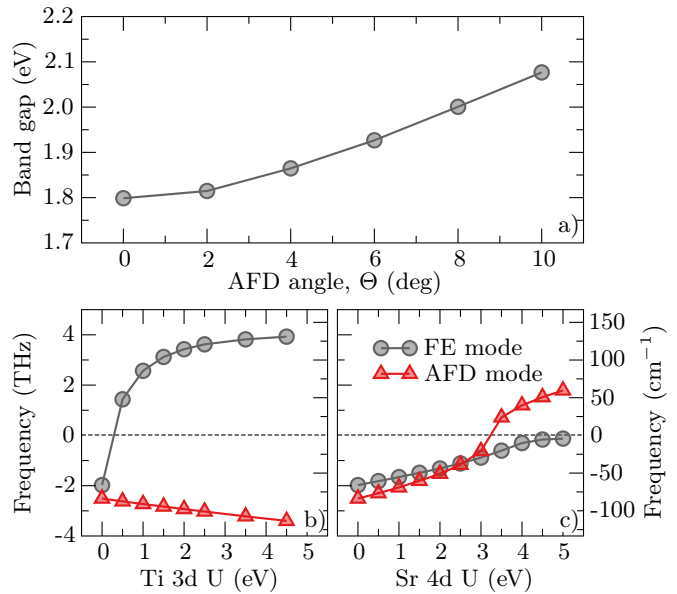
\caption{\label{fig:sojt}a) Evolution of the PBEsol band-gap as a function of the AFD angle. The effect on the FE and AFD phonon frequencies of artificially increasing the energy of b) Ti 3d and c) Sr 4d conduction-band states by an additional on-site energy U.} 
\end{figure}

The observed suppression of the FE distortion by AFD rotations can also be reconciled with a second order Jahn-Teller picture. In this formalism the energy as a function of a distortion along a normal-coordinate $Q$ is expressed as a Taylor expansion around the unperturbed energy $E^0$ using standard perturbation theory \cite{Rondinelli:2009tq}:
\begin{multline}
E=E^0+\left<0\left|H^1\right|0\right>Q \\ +\frac{1}{2}\left[ \left<0\left|H^2\right|0\right>  -2\sum_{n}{\frac{\left|\left<0\left|H^1\right|n\right>\right|^2}{E^n-E^0}}  \right]Q^2+...
\end{multline}
Where $H^1$ and $H^2$ respectively are the first and second derivative of the ground-state Hamiltonian with respect to $Q$, $\left|0\right>$ is the ground-state solution of the unperturbed Hamiltonian with energy $E^0$ and $\left|n\right>$ are excited states with energies $E^n$. The first-order term represents the usual Jahn-Teller distortion for orbitally-degenerate systems. The first of the second-order terms is always positive and represents the increase in energy due to repulsive forces when ions are displaced while electrons are kept frozen. It always favours the centro-symmetric structure. The second of the second-order terms represents the energy gain due to hybridisation following the distortion; ferroelectricity results if it dominates over the preceding term for a polar distortion Q. This term can only be non-zero for states of opposite parity such as those formed from the O$_\mathrm{2p}$ and Ti$_\mathrm{3d}$ orbitals that make up the valence and conduction bands respectively in SrTiO$_3$. The tendency for ferroelectricity is larger for smaller energy differences between these states, which for the case of STO means a smaller band gap. It is hence instructive to follow the evolution of the band gap as a function of the AFD angle, which we show in Fig. \ref{fig:sojt}a). We see that the AFD rotation leads to an opening of the band-gap \cite{Berger:2011jv} and hence weakens the FE instability in STO by reducing the tendency for cross-gap hybridisation.

In this context it is interesting to consider if the underestimation of the band-gap within PBEsol is at the origin of the overestimated equilibrium AFD angle compared to experiment. In contrast, hybrid functionals, which yield a better description of the band-gap, also result in correct AFD angles. We have verified that the effect of a HSE hybrid-functional on the band-structure can, as a first approximation, be described by a rigid upwards shift of the conduction band relative to the top of the valence band. In order to test the hypothesis that this will affect the AFD instability we performed a computer experiment within PBEsol by applying an additional on-site energy to either the Ti 3d and Sr 4d states by means of the Hubbard U correction. The resulting dependency of the phonon frequencies in the cubic structure on the value of U are shown in Figs. \ref{fig:sojt}b) and c) respectively. We see that the artificial increase in the energy of Ti 3d states at the bottom of the conduction band results in a very rapid suppression of the FE mode. This is in line with the small increase in band-gap shown in \ref{fig:sojt}a) leading to the marked reduction of the FE instability. The AFD instability on the other hand is strengthened when the Ti states are shifted up, which hints at the fact that the coordination is increasingly optimised via Sr-O bond formation. Applying an on-site energy to the Sr 4d states supports this picture in the sense that both the AFD and the FE modes are rendered less unstable with increasing U. AFD rotations are no longer energy lowering beyond U $\sim$3.25 eV, whereas a small FE instability persists up to the highest U values tried here. This indicates that the AFD vanishes when Sr-O hybridisation is reduced, while the FE still persists, owing to its Ti-O hybridisation component. From Figs. \ref{fig:sojt}b) and c) we see that the suppression of the AFD from shifting the Sr states is more marked than the enhancement from shifting the Ti states, resulting in a net suppression of the AFD as the whole conduction band is shifted upwards. This offers an explanation as to why hybrid functionals describe AFD rotation angles better than (semi-)local functionals.

\section{Conclusions}

Using density functional theory calculations with the PBEsol functional, we have shown that a competition exists between the antiferrodistortive (AFD) and the ferroelectric (FE) instability in SrTiO$_3$. The AFD reduces the energy gain due to the FE by a factor of five between the cubic structure and that with equilibrium AFD rotations.  The suppression of the FE over this range stems from charge transfer from Ti-O $\sigma^*$ bonds to Sr-O $\sigma$ bonds. This decrease of anti-bonding overlap causes a strengthening of the Ti-O bonds and opposes Ti off-centering. At larger AFD angles we observe a surprising reappearance of the FE instability, which is driven by the increased tetragonality of the cell.

We further show by analysing interatomic force constants that differences between (semi)-local density functionals are mostly due to their effect on the Ti-O interactions and this is why they have a strong effect on the FE instability. We also show that the underestimation of the band-gap using PBEsol is at the origin of its overestimated AFD angle, as the AFD instability is strongly affected by the energy of the empty Sr 4d states.

\section{Acknowledgements}

This work was supported financially by the ETH Z\"urich and by the ERC Advanced Grant program, No. 291151. We are grateful to Dr. Michael Fechner and Dr. Eric Bousquet for useful discussions.

\bibliography{references.bib}

\end{document}

%% file: figure1.pdf_tex
\begingroup%
  \makeatletter%
  \providecommand\color[2][]{%
    \errmessage{(Inkscape) Color is used for the text in Inkscape, but the package 'color.sty' is not loaded}%
    \renewcommand\color[2][]{}%
  }%
  \providecommand\transparent[1]{%
    \errmessage{(Inkscape) Transparency is used (non-zero) for the text in Inkscape, but the package 'transparent.sty' is not loaded}%
    \renewcommand\transparent[1]{}%
  }%
  \providecommand\rotatebox[2]{#2}%
  \ifx\svgwidth\undefined%
    \setlength{\unitlength}{226.77165222bp}%
    \ifx\svgscale\undefined%
      \relax%
    \else%
      \setlength{\unitlength}{\unitlength * \real{\svgscale}}%
    \fi%
  \else%
    \setlength{\unitlength}{\svgwidth}%
  \fi%
  \global\let\svgwidth\undefined%
  \global\let\svgscale\undefined%
  \makeatother%
  \begin{picture}(1,0.83560557)%
    \put(0,0){\includegraphics[width=\unitlength]{figure1.pdf}}%
    \put(0.29176416,0.62775892){\color[rgb]{0,0,0}\makebox(0,0)[lb]{\smash{b}}}%
    \put(0.13138037,0.79475298){\color[rgb]{0,0,0}\makebox(0,0)[lb]{\smash{c}}}%
    \put(0.16235455,0.57561207){\color[rgb]{0,0,0}\makebox(0,0)[lb]{\smash{a}}}%
    \put(0.11394011,0.00754476){\color[rgb]{0,0,0}\makebox(0,0)[lb]{\smash{a) Polar FE mode}}}%
    \put(0.56635851,0.0076761){\color[rgb]{0,0,0}\makebox(0,0)[lb]{\smash{b) Rotational AFD mode}}}%
  \end{picture}%
\endgroup%

%% file: figure2.pdf_tex

\begingroup
  \makeatletter
  \providecommand\color[2][]{%
    \errmessage{(Inkscape) Color is used for the text in Inkscape, but the package 'color.sty' is not loaded}
    \renewcommand\color[2][]{}%
  }
  \providecommand\transparent[1]{%
    \errmessage{(Inkscape) Transparency is used (non-zero) for the text in Inkscape, but the package 'transparent.sty' is not loaded}
    \renewcommand\transparent[1]{}%
  }
  \providecommand\rotatebox[2]{#2}
  \ifx\svgwidth\undefined
    \setlength{\unitlength}{228.34766769pt}
  \else
    \setlength{\unitlength}{\svgwidth}
  \fi
  \global\let\svgwidth\undefined
  \makeatother
  \begin{picture}(1,0.79441642)%
    \put(0,0){\includegraphics[width=\unitlength]{figure2.pdf}}%
    \put(0.08336768,0.44481906){\color[rgb]{0,0,0}\makebox(0,0)[lb]{\smash{R}}}%
    \put(0.36806972,0.44534936){\color[rgb]{0,0,0}\makebox(0,0)[lb]{\smash{$\Gamma$}}}%
    \put(0.52969974,0.44481906){\color[rgb]{0,0,0}\makebox(0,0)[lb]{\smash{X}}}%
    \put(0.7266562,0.44481906){\color[rgb]{0,0,0}\makebox(0,0)[lb]{\smash{M}}}%
    \put(0.85828085,0.44481906){\color[rgb]{0,0,0}\makebox(0,0)[lb]{\smash{R}}}%
    \put(0.06338222,0.50430923){\color[rgb]{0,0,0}\makebox(0,0)[lb]{\smash{0}}}%
    \put(0.06444283,0.55674414){\color[rgb]{0,0,0}\makebox(0,0)[lb]{\smash{5}}}%
    \put(0.04586507,0.61130775){\color[rgb]{0,0,0}\makebox(0,0)[lb]{\smash{10}}}%
    \put(0.04692568,0.66374252){\color[rgb]{0,0,0}\makebox(0,0)[lb]{\smash{15}}}%
    \put(0.04586507,0.71830586){\color[rgb]{0,0,0}\makebox(0,0)[lb]{\smash{20}}}%
    \put(0.04692568,0.7707409){\color[rgb]{0,0,0}\makebox(0,0)[lb]{\smash{25}}}%
    \put(0.02432557,0.49832687){\color[rgb]{0,0,0}\rotatebox{90}{\makebox(0,0)[lb]{\smash{Frequency (THz)}}}}%
    \put(0.8822945,0.50430923){\color[rgb]{0,0,0}\makebox(0,0)[lb]{\smash{0}}}%
    \put(0.88280769,0.56846563){\color[rgb]{0,0,0}\makebox(0,0)[lb]{\smash{200}}}%
    \put(0.88301297,0.63262162){\color[rgb]{0,0,0}\makebox(0,0)[lb]{\smash{400}}}%
    \put(0.882055,0.69572436){\color[rgb]{0,0,0}\makebox(0,0)[lb]{\smash{600}}}%
    \put(0.88143917,0.76092332){\color[rgb]{0,0,0}\makebox(0,0)[lb]{\smash{800}}}%
    \put(0.99243889,0.48319556){\color[rgb]{0,0,0}\rotatebox{90}{\makebox(0,0)[lb]{\smash{Frequency (cm-1)}}}}%
    \put(0.06421198,0.0545731){\color[rgb]{0,0,0}\makebox(0,0)[lb]{\smash{3.86}}}%
    \put(0.23747908,0.0545731){\color[rgb]{0,0,0}\makebox(0,0)[lb]{\smash{3.88}}}%
    \put(0.41713142,0.0545731){\color[rgb]{0,0,0}\makebox(0,0)[lb]{\smash{3.9}}}%
    \put(0.58082071,0.0545731){\color[rgb]{0,0,0}\makebox(0,0)[lb]{\smash{3.92}}}%
    \put(0.75302367,0.0545731){\color[rgb]{0,0,0}\makebox(0,0)[lb]{\smash{3.94}}}%
    \put(0.31101437,0.00749269){\color[rgb]{0,0,0}\makebox(0,0)[lb]{\smash{Lattice parameter (\AA)}}}%
    \put(0.0516471,0.11472201){\color[rgb]{0,0,0}\makebox(0,0)[lb]{\smash{-4}}}%
    \put(0.05188659,0.18129872){\color[rgb]{0,0,0}\makebox(0,0)[lb]{\smash{-2}}}%
    \put(0.06338222,0.24787543){\color[rgb]{0,0,0}\makebox(0,0)[lb]{\smash{0}}}%
    \put(0.0635875,0.31445201){\color[rgb]{0,0,0}\makebox(0,0)[lb]{\smash{2}}}%
    \put(0.06334801,0.38102872){\color[rgb]{0,0,0}\makebox(0,0)[lb]{\smash{4}}}%
    \put(0.03071079,0.10755074){\color[rgb]{0,0,0}\rotatebox{90}{\makebox(0,0)[lb]{\smash{Frequency (THz)}}}}%
    \put(0.88214054,0.09716343){\color[rgb]{0,0,0}\makebox(0,0)[lb]{\smash{-150}}}%
    \put(0.88214054,0.1481102){\color[rgb]{0,0,0}\makebox(0,0)[lb]{\smash{-100}}}%
    \put(0.88214054,0.19692866){\color[rgb]{0,0,0}\makebox(0,0)[lb]{\smash{-50}}}%
    \put(0.8822945,0.24787543){\color[rgb]{0,0,0}\makebox(0,0)[lb]{\smash{0}}}%
    \put(0.88188394,0.29669363){\color[rgb]{0,0,0}\makebox(0,0)[lb]{\smash{50}}}%
    \put(0.8794548,0.34764053){\color[rgb]{0,0,0}\makebox(0,0)[lb]{\smash{100}}}%
    \put(0.99243889,0.09241916){\color[rgb]{0,0,0}\rotatebox{90}{\makebox(0,0)[lb]{\smash{Frequency (cm$^{-1}$)}}}}%
    \put(0.63242956,0.33204961){\color[rgb]{0,0,0}\makebox(0,0)[lb]{\smash{LDA}}}%
    \put(0.63242956,0.29561124){\color[rgb]{0,0,0}\makebox(0,0)[lb]{\smash{PBE}}}%
    \put(0.63242956,0.25704443){\color[rgb]{0,0,0}\makebox(0,0)[lb]{\smash{PBEsol}}}%
    \put(0.63242956,0.22060605){\color[rgb]{0,0,0}\makebox(0,0)[lb]{\smash{Experiment}}}%
    \put(0.82661651,0.35932589){\color[rgb]{0,0,0}\makebox(0,0)[lb]{\smash{b)}}}%
    \put(0.82661651,0.74392786){\color[rgb]{0,0,0}\makebox(0,0)[lb]{\smash{a)}}}%
  \end{picture}%
\endgroup

%% file: figure3.pdf_tex
\begingroup%
  \makeatletter%
  \providecommand\color[2][]{%
    \errmessage{(Inkscape) Color is used for the text in Inkscape, but the package 'color.sty' is not loaded}%
    \renewcommand\color[2][]{}%
  }%
  \providecommand\transparent[1]{%
    \errmessage{(Inkscape) Transparency is used (non-zero) for the text in Inkscape, but the package 'transparent.sty' is not loaded}%
    \renewcommand\transparent[1]{}%
  }%
  \providecommand\rotatebox[2]{#2}%
  \ifx\svgwidth\undefined%
    \setlength{\unitlength}{226.68670484bp}%
    \ifx\svgscale\undefined%
      \relax%
    \else%
      \setlength{\unitlength}{\unitlength * \real{\svgscale}}%
    \fi%
  \else%
    \setlength{\unitlength}{\svgwidth}%
  \fi%
  \global\let\svgwidth\undefined%
  \global\let\svgscale\undefined%
  \makeatother%
  \begin{picture}(1,0.43303339)%
    \put(0,0){\includegraphics[width=\unitlength]{figure3.pdf}}%
    \put(0.18900164,0.05347823){\color[rgb]{0,0,0}\makebox(0,0)[lb]{\smash{0}}}%
    \put(0.33455038,0.05347823){\color[rgb]{0,0,0}\makebox(0,0)[lb]{\smash{2}}}%
    \put(0.48009909,0.05347823){\color[rgb]{0,0,0}\makebox(0,0)[lb]{\smash{4}}}%
    \put(0.62564773,0.05107949){\color[rgb]{0,0,0}\makebox(0,0)[lb]{\smash{6}}}%
    \put(0.77239588,0.05347823){\color[rgb]{0,0,0}\makebox(0,0)[lb]{\smash{8}}}%
    \put(0.90595143,0.05347823){\color[rgb]{0,0,0}\makebox(0,0)[lb]{\smash{10}}}%
    \put(0.39622434,0.00761651){\color[rgb]{0,0,0}\makebox(0,0)[lb]{\smash{AFD angle, $\Theta$ (deg)}}}%
    \put(0.04637926,0.12235466){\color[rgb]{0,0,0}\makebox(0,0)[lb]{\smash{-0.6}}}%
    \put(0.04637926,0.20796142){\color[rgb]{0,0,0}\makebox(0,0)[lb]{\smash{-0.4}}}%
    \put(0.04637926,0.29236908){\color[rgb]{0,0,0}\makebox(0,0)[lb]{\smash{-0.2}}}%
    \put(0.060771,0.37677661){\color[rgb]{0,0,0}\makebox(0,0)[lb]{\smash{0.0}}}%
    \put(0.02450381,0.09902263){\color[rgb]{0,0,0}\rotatebox{90}{\makebox(0,0)[lb]{\smash{E$_\mathrm{FE}$-E$_\mathrm{PE}$ (meV/f.u.)}}}}%
  \end{picture}%
\endgroup%

%% file: figure4.pdf_tex

\begingroup
  \makeatletter
  \providecommand\color[2][]{%
    \errmessage{(Inkscape) Color is used for the text in Inkscape, but the package 'color.sty' is not loaded}
    \renewcommand\color[2][]{}%
  }
  \providecommand\transparent[1]{%
    \errmessage{(Inkscape) Transparency is used (non-zero) for the text in Inkscape, but the package 'transparent.sty' is not loaded}
    \renewcommand\transparent[1]{}%
  }
  \providecommand\rotatebox[2]{#2}
  \ifx\svgwidth\undefined
    \setlength{\unitlength}{240.5510273pt}
  \else
    \setlength{\unitlength}{\svgwidth}
  \fi
  \global\let\svgwidth\undefined
  \makeatother
  \begin{picture}(1,1.22858172)%
    \put(0,0){\includegraphics[width=\unitlength]{figure4.pdf}}%
    \put(0.06364849,0.87969693){\color[rgb]{0,0,0}\makebox(0,0)[lb]{\smash{3.84}}}%
    \put(0.0637784,0.9404172){\color[rgb]{0,0,0}\makebox(0,0)[lb]{\smash{3.86}}}%
    \put(0.06441171,1.00284786){\color[rgb]{0,0,0}\makebox(0,0)[lb]{\smash{3.88}}}%
    \put(0.06368096,1.06442333){\color[rgb]{0,0,0}\makebox(0,0)[lb]{\smash{3.90}}}%
    \put(0.06387583,1.1259988){\color[rgb]{0,0,0}\makebox(0,0)[lb]{\smash{3.92}}}%
    \put(0.06364849,1.18757414){\color[rgb]{0,0,0}\makebox(0,0)[lb]{\smash{3.94}}}%
    \put(0.02309151,0.86326077){\color[rgb]{0,0,0}\rotatebox{90}{\makebox(0,0)[lb]{\smash{Lattice parameter (\AA)}}}}%
    \put(0.26629134,0.99543525){\color[rgb]{0,0,0}\makebox(0,0)[lb]{\smash{a}}}%
    \put(0.26634006,0.95376064){\color[rgb]{0,0,0}\makebox(0,0)[lb]{\smash{c}}}%
    \put(0.26629134,0.91339514){\color[rgb]{0,0,0}\makebox(0,0)[lb]{\smash{a$_0$ $\cdot$ cos($\Theta$)}}}%
    \put(0.04705246,0.4979293){\color[rgb]{0,0,0}\makebox(0,0)[lb]{\smash{1.000}}}%
    \put(0.04805926,0.57096464){\color[rgb]{0,0,0}\makebox(0,0)[lb]{\smash{1.005}}}%
    \put(0.04705246,0.64571037){\color[rgb]{0,0,0}\makebox(0,0)[lb]{\smash{1.010}}}%
    \put(0.04805926,0.71874546){\color[rgb]{0,0,0}\makebox(0,0)[lb]{\smash{1.015}}}%
    \put(0.04705246,0.79349145){\color[rgb]{0,0,0}\makebox(0,0)[lb]{\smash{1.020}}}%
    \put(0.02309151,0.62670519){\color[rgb]{0,0,0}\rotatebox{90}{\makebox(0,0)[lb]{\smash{c/a}}}}%
    \put(0.26392547,0.67785698){\color[rgb]{0,0,0}\makebox(0,0)[lb]{\smash{c/a}}}%
    \put(0.26476988,0.63588892){\color[rgb]{0,0,0}\makebox(0,0)[lb]{\smash{Volume}}}%
    \put(0.19669137,0.04653216){\color[rgb]{0,0,0}\makebox(0,0)[lb]{\smash{0}}}%
    \put(0.31881599,0.04653216){\color[rgb]{0,0,0}\makebox(0,0)[lb]{\smash{2}}}%
    \put(0.44094062,0.04653216){\color[rgb]{0,0,0}\makebox(0,0)[lb]{\smash{4}}}%
    \put(0.56306532,0.04482176){\color[rgb]{0,0,0}\makebox(0,0)[lb]{\smash{6}}}%
    \put(0.68604511,0.04653216){\color[rgb]{0,0,0}\makebox(0,0)[lb]{\smash{8}}}%
    \put(0.79961759,0.04653216){\color[rgb]{0,0,0}\makebox(0,0)[lb]{\smash{10}}}%
    \put(0.34398109,0.00717753){\color[rgb]{0,0,0}\makebox(0,0)[lb]{\smash{AFD angle, $\Theta$ (deg)}}}%
    \put(0.04705246,0.10299103){\color[rgb]{0,0,0}\makebox(0,0)[lb]{\smash{1.950}}}%
    \put(0.04805926,0.16456662){\color[rgb]{0,0,0}\makebox(0,0)[lb]{\smash{1.955}}}%
    \put(0.04705246,0.22614196){\color[rgb]{0,0,0}\makebox(0,0)[lb]{\smash{1.960}}}%
    \put(0.04805926,0.28771756){\color[rgb]{0,0,0}\makebox(0,0)[lb]{\smash{1.965}}}%
    \put(0.04705246,0.35014823){\color[rgb]{0,0,0}\makebox(0,0)[lb]{\smash{1.970}}}%
    \put(0.04805926,0.41086824){\color[rgb]{0,0,0}\makebox(0,0)[lb]{\smash{1.975}}}%
    \put(0.02309151,0.08912059){\color[rgb]{0,0,0}\rotatebox{90}{\makebox(0,0)[lb]{\smash{Ti-O bond length (\AA)}}}}%
    \put(0.26476988,0.31416568){\color[rgb]{0,0,0}\makebox(0,0)[lb]{\smash{In plane}}}%
    \put(0.26442887,0.27300702){\color[rgb]{0,0,0}\makebox(0,0)[lb]{\smash{Out-of-plane}}}%
    \put(0.88568637,0.49707397){\color[rgb]{0,0,0}\makebox(0,0)[lb]{\smash{58.9}}}%
    \put(0.88568637,0.57096464){\color[rgb]{0,0,0}\makebox(0,0)[lb]{\smash{59.0}}}%
    \put(0.88568637,0.64485505){\color[rgb]{0,0,0}\makebox(0,0)[lb]{\smash{59.1}}}%
    \put(0.88568637,0.71874546){\color[rgb]{0,0,0}\makebox(0,0)[lb]{\smash{59.2}}}%
    \put(0.88568637,0.79263612){\color[rgb]{0,0,0}\makebox(0,0)[lb]{\smash{59.3}}}%
    \put(0.99288742,0.55068801){\color[rgb]{0,0,0}\rotatebox{90}{\makebox(0,0)[lb]{\smash{Volume (\AA$^3$)}}}}%
    \put(0.83164404,0.4963511){\color[rgb]{0,0,0}\makebox(0,0)[lb]{\smash{b)}}}%
    \put(0.83126179,0.10840158){\color[rgb]{0,0,0}\makebox(0,0)[lb]{\smash{c)}}}%
    \put(0.83202622,0.88391774){\color[rgb]{0,0,0}\makebox(0,0)[lb]{\smash{a)}}}%
  \end{picture}%
\endgroup

%% file: figure5.pdf_tex
\begingroup%
  \makeatletter%
  \providecommand\color[2][]{%
    \errmessage{(Inkscape) Color is used for the text in Inkscape, but the package 'color.sty' is not loaded}%
    \renewcommand\color[2][]{}%
  }%
  \providecommand\transparent[1]{%
    \errmessage{(Inkscape) Transparency is used (non-zero) for the text in Inkscape, but the package 'transparent.sty' is not loaded}%
    \renewcommand\transparent[1]{}%
  }%
  \providecommand\rotatebox[2]{#2}%
  \ifx\svgwidth\undefined%
    \setlength{\unitlength}{222.50352097bp}%
    \ifx\svgscale\undefined%
      \relax%
    \else%
      \setlength{\unitlength}{\unitlength * \real{\svgscale}}%
    \fi%
  \else%
    \setlength{\unitlength}{\svgwidth}%
  \fi%
  \global\let\svgwidth\undefined%
  \global\let\svgscale\undefined%
  \makeatother%
  \begin{picture}(1,1.27394348)%
    \put(0,0){\includegraphics[width=\unitlength]{figure5.pdf}}%
    \put(0.15870595,0.71600032){\color[rgb]{0,0,0}\makebox(0,0)[lb]{\smash{0}}}%
    \put(0.30997127,0.71600032){\color[rgb]{0,0,0}\makebox(0,0)[lb]{\smash{2}}}%
    \put(0.46123659,0.71600032){\color[rgb]{0,0,0}\makebox(0,0)[lb]{\smash{4}}}%
    \put(0.61250195,0.71600032){\color[rgb]{0,0,0}\makebox(0,0)[lb]{\smash{6}}}%
    \put(0.76482655,0.71600032){\color[rgb]{0,0,0}\makebox(0,0)[lb]{\smash{8}}}%
    \put(0.95385321,0.57095551){\color[rgb]{0,0,0}\makebox(0,0)[lb]{\smash{b)}}}%
    \put(0.95463322,1.23606662){\color[rgb]{0,0,0}\makebox(0,0)[lb]{\smash{a)}}}%
    \put(0.90549905,0.71600032){\color[rgb]{0,0,0}\makebox(0,0)[lb]{\smash{10}}}%
    \put(0.29207344,0.67042094){\color[rgb]{0,0,0}\makebox(0,0)[lb]{\smash{Equivalent ADF angle, $\Theta$ (deg)}}}%
    \put(0.04442759,0.13368655){\color[rgb]{0,0,0}\makebox(0,0)[lb]{\smash{-4}}}%
    \put(0.04442759,0.79459891){\color[rgb]{0,0,0}\makebox(0,0)[lb]{\smash{-4}}}%
    \put(0.04616563,0.23741121){\color[rgb]{0,0,0}\makebox(0,0)[lb]{\smash{-3}}}%
    \put(0.04616563,0.89832371){\color[rgb]{0,0,0}\makebox(0,0)[lb]{\smash{-3}}}%
    \put(0.04467338,0.34113614){\color[rgb]{0,0,0}\makebox(0,0)[lb]{\smash{-2}}}%
    \put(0.04467338,1.0020485){\color[rgb]{0,0,0}\makebox(0,0)[lb]{\smash{-2}}}%
    \put(0.0475701,0.4448608){\color[rgb]{0,0,0}\makebox(0,0)[lb]{\smash{-1}}}%
    \put(0.0475701,1.1057733){\color[rgb]{0,0,0}\makebox(0,0)[lb]{\smash{-1}}}%
    \put(0.05647094,0.54858573){\color[rgb]{0,0,0}\makebox(0,0)[lb]{\smash{0}}}%
    \put(0.05647094,1.20949809){\color[rgb]{0,0,0}\makebox(0,0)[lb]{\smash{0}}}%
    \put(0.15870595,0.05333867){\color[rgb]{0,0,0}\makebox(0,0)[lb]{\smash{0}}}%
    \put(0.30997127,0.05333867){\color[rgb]{0,0,0}\makebox(0,0)[lb]{\smash{2}}}%
    \put(0.46123659,0.05333867){\color[rgb]{0,0,0}\makebox(0,0)[lb]{\smash{4}}}%
    \put(0.61250195,0.05333867){\color[rgb]{0,0,0}\makebox(0,0)[lb]{\smash{6}}}%
    \put(0.76482655,0.05333867){\color[rgb]{0,0,0}\makebox(0,0)[lb]{\smash{8}}}%
    \put(0.90549905,0.05333867){\color[rgb]{0,0,0}\makebox(0,0)[lb]{\smash{10}}}%
    \put(0.37575636,0.00775971){\color[rgb]{0,0,0}\makebox(0,0)[lb]{\smash{AFD angle, $\Theta$ (deg)}}}%
    \put(0.02496449,0.83538018){\color[rgb]{0,0,0}\rotatebox{90}{\makebox(0,0)[lb]{\smash{E$_\mathrm{FE}$ - E$_\mathrm{PE}$ (meV/f.u.)}}}}%
    \put(0.02496449,0.17272018){\color[rgb]{0,0,0}\rotatebox{90}{\makebox(0,0)[lb]{\smash{E$_\mathrm{FE}$ - E$_\mathrm{PE}$ (meV/f.u.)}}}}%
    \put(0.25495587,0.24451433){\color[rgb]{0,0,0}\makebox(0,0)[lt]{\begin{minipage}{0.61875568\unitlength}\raggedright Ideal tetragonal\\ a=b=a$_0$$\cdot$cos($\Theta$)\\ c=a$_0$/cos$^2$($\Theta$)\end{minipage}}}%
    \put(0.25495587,0.38471176){\color[rgb]{0,0,0}\makebox(0,0)[lt]{\begin{minipage}{0.47996744\unitlength}\raggedright In plane only\\ a=b=a$_0$$\cdot$cos($\Theta$)\\ c=const\end{minipage}}}%
    \put(0.27883994,1.01800944){\color[rgb]{0,0,0}\makebox(0,0)[lt]{\begin{minipage}{0.34479284\unitlength}\raggedright Ti-O bonds\\ Fig. 4c\end{minipage}}}%
    \put(0.27883994,0.90818509){\color[rgb]{0,0,0}\makebox(0,0)[lt]{\begin{minipage}{0.34479284\unitlength}\raggedright Tetragonal\\ Fig. 4a and\\ c=const=a$_0$\end{minipage}}}%
    \put(0.27883994,1.12783377){\color[rgb]{0,0,0}\makebox(0,0)[lt]{\begin{minipage}{0.34608992\unitlength}\raggedright Tetragonal\\ Fig. 4a\end{minipage}}}%
    \put(0.25456962,0.48895465){\color[rgb]{0,0,0}\makebox(0,0)[lt]{\begin{minipage}{0.36554622\unitlength}\raggedright Cubic cell\\ a=b=c=a$_0$\end{minipage}}}%
  \end{picture}%
\endgroup%

%% file: figure6.pdf_tex
\begingroup%
  \makeatletter%
  \providecommand\color[2][]{%
    \errmessage{(Inkscape) Color is used for the text in Inkscape, but the package 'color.sty' is not loaded}%
    \renewcommand\color[2][]{}%
  }%
  \providecommand\transparent[1]{%
    \errmessage{(Inkscape) Transparency is used (non-zero) for the text in Inkscape, but the package 'transparent.sty' is not loaded}%
    \renewcommand\transparent[1]{}%
  }%
  \providecommand\rotatebox[2]{#2}%
  \ifx\svgwidth\undefined%
    \setlength{\unitlength}{228.79625893bp}%
    \ifx\svgscale\undefined%
      \relax%
    \else%
      \setlength{\unitlength}{\unitlength * \real{\svgscale}}%
    \fi%
  \else%
    \setlength{\unitlength}{\svgwidth}%
  \fi%
  \global\let\svgwidth\undefined%
  \global\let\svgscale\undefined%
  \makeatother%
  \begin{picture}(1,0.79800435)%
    \put(0,0){\includegraphics[width=\unitlength]{figure6.pdf}}%
    \put(0.04402679,0.47300952){\color[rgb]{0,0,0}\makebox(0,0)[lb]{\smash{-0.04}}}%
    \put(0.04613778,0.52954413){\color[rgb]{0,0,0}\makebox(0,0)[lb]{\smash{-0.02}}}%
    \put(0.05597738,0.58687023){\color[rgb]{0,0,0}\makebox(0,0)[lb]{\smash{0.00}}}%
    \put(0.0569892,0.64419633){\color[rgb]{0,0,0}\makebox(0,0)[lb]{\smash{0.02}}}%
    \put(0.05566943,0.70152216){\color[rgb]{0,0,0}\makebox(0,0)[lb]{\smash{0.04}}}%
    \put(0.05587473,0.75797193){\color[rgb]{0,0,0}\makebox(0,0)[lb]{\smash{0.06}}}%
    \put(0.02427788,0.49033106){\color[rgb]{0,0,0}\rotatebox{90}{\makebox(0,0)[lb]{\smash{Ionic displ. (\AA)}}}}%
    \put(0.15273155,0.05328872){\color[rgb]{0,0,0}\makebox(0,0)[lb]{\smash{0}}}%
    \put(0.2137721,0.05328872){\color[rgb]{0,0,0}\makebox(0,0)[lb]{\smash{2}}}%
    \put(0.27481293,0.05328872){\color[rgb]{0,0,0}\makebox(0,0)[lb]{\smash{4}}}%
    \put(0.33585355,0.05328872){\color[rgb]{0,0,0}\makebox(0,0)[lb]{\smash{6}}}%
    \put(0.39689417,0.05328872){\color[rgb]{0,0,0}\makebox(0,0)[lb]{\smash{8}}}%
    \put(0.45004858,0.05328872){\color[rgb]{0,0,0}\makebox(0,0)[lb]{\smash{10}}}%
    \put(0.15653644,0.00754629){\color[rgb]{0,0,0}\makebox(0,0)[lb]{\smash{AFD angle, $\Theta$ (deg)}}}%
    \put(0.04481801,0.11106339){\color[rgb]{0,0,0}\makebox(0,0)[lb]{\smash{-0.04}}}%
    \put(0.04613778,0.16838949){\color[rgb]{0,0,0}\makebox(0,0)[lb]{\smash{-0.02}}}%
    \put(0.05597738,0.22571559){\color[rgb]{0,0,0}\makebox(0,0)[lb]{\smash{0.00}}}%
    \put(0.0569892,0.28304143){\color[rgb]{0,0,0}\makebox(0,0)[lb]{\smash{0.02}}}%
    \put(0.05566943,0.34036753){\color[rgb]{0,0,0}\makebox(0,0)[lb]{\smash{0.04}}}%
    \put(0.05587473,0.3968173){\color[rgb]{0,0,0}\makebox(0,0)[lb]{\smash{0.06}}}%
    \put(0.02427788,0.13616918){\color[rgb]{0,0,0}\rotatebox{90}{\makebox(0,0)[lb]{\smash{Ionic displ. (\AA)}}}}%
    \put(0.88983361,0.48567346){\color[rgb]{0,0,0}\makebox(0,0)[lb]{\smash{-0.6}}}%
    \put(0.88962833,0.57253873){\color[rgb]{0,0,0}\makebox(0,0)[lb]{\smash{-0.4}}}%
    \put(0.89094816,0.65852782){\color[rgb]{0,0,0}\makebox(0,0)[lb]{\smash{-0.2}}}%
    \put(0.90078771,0.74451677){\color[rgb]{0,0,0}\makebox(0,0)[lb]{\smash{0.0}}}%
    \put(0.992522,0.4591314){\color[rgb]{0,0,0}\rotatebox{90}{\makebox(0,0)[lb]{\smash{E$_\mathrm{FE}$-E$_\mathrm{PE}$ (meV/f.u.)}}}}%
    \put(0.5372892,0.05328872){\color[rgb]{0,0,0}\makebox(0,0)[lb]{\smash{0}}}%
    \put(0.59832982,0.05328872){\color[rgb]{0,0,0}\makebox(0,0)[lb]{\smash{2}}}%
    \put(0.65937037,0.05328872){\color[rgb]{0,0,0}\makebox(0,0)[lb]{\smash{4}}}%
    \put(0.720411,0.05328872){\color[rgb]{0,0,0}\makebox(0,0)[lb]{\smash{6}}}%
    \put(0.78145182,0.05328872){\color[rgb]{0,0,0}\makebox(0,0)[lb]{\smash{8}}}%
    \put(0.83460589,0.05328872){\color[rgb]{0,0,0}\makebox(0,0)[lb]{\smash{10}}}%
    \put(0.54109362,0.00754629){\color[rgb]{0,0,0}\makebox(0,0)[lb]{\smash{AFD angle, $\Theta$ (deg)}}}%
    \put(0.88983361,0.12451882){\color[rgb]{0,0,0}\makebox(0,0)[lb]{\smash{-0.6}}}%
    \put(0.88962833,0.2113841){\color[rgb]{0,0,0}\makebox(0,0)[lb]{\smash{-0.4}}}%
    \put(0.89094816,0.29737305){\color[rgb]{0,0,0}\makebox(0,0)[lb]{\smash{-0.2}}}%
    \put(0.90078771,0.38336214){\color[rgb]{0,0,0}\makebox(0,0)[lb]{\smash{0.0}}}%
    \put(0.992522,0.09797717){\color[rgb]{0,0,0}\rotatebox{90}{\makebox(0,0)[lb]{\smash{E$_\mathrm{FE}$-E$_\mathrm{PE}$ (meV/f.u.)}}}}%
    \put(0.46005405,0.47369178){\color[rgb]{0,0,0}\makebox(0,0)[lb]{\smash{a)}}}%
    \put(0.84300217,0.47369178){\color[rgb]{0,0,0}\makebox(0,0)[lb]{\smash{b)}}}%
    \put(0.46005405,0.11044956){\color[rgb]{0,0,0}\makebox(0,0)[lb]{\smash{c)}}}%
    \put(0.84300217,0.11044956){\color[rgb]{0,0,0}\makebox(0,0)[lb]{\smash{d)}}}%
    \put(0.23889726,0.50837741){\color[rgb]{0,0,0}\makebox(0,0)[lb]{\smash{Sr}}}%
    \put(0.23889726,0.47471016){\color[rgb]{0,0,0}\makebox(0,0)[lb]{\smash{Ti}}}%
    \put(0.23731481,0.75752857){\color[rgb]{0,0,0}\makebox(0,0)[lb]{\smash{O$_\mathrm{apical}$}}}%
    \put(0.23731481,0.71599517){\color[rgb]{0,0,0}\makebox(0,0)[lb]{\smash{O$_\mathrm{equatorial}$}}}%
    \put(0.4017112,0.75008752){\color[rgb]{0,0,0}\makebox(0,0)[lb]{\smash{Fully relaxed}}}%
    \put(0.34869922,0.3890556){\color[rgb]{0,0,0}\makebox(0,0)[lb]{\smash{Sr displ. set to zero}}}%
  \end{picture}%
\endgroup%

%% file: figure7.pdf_tex
\begingroup%
  \makeatletter%
  \providecommand\color[2][]{%
    \errmessage{(Inkscape) Color is used for the text in Inkscape, but the package 'color.sty' is not loaded}%
    \renewcommand\color[2][]{}%
  }%
  \providecommand\transparent[1]{%
    \errmessage{(Inkscape) Transparency is used (non-zero) for the text in Inkscape, but the package 'transparent.sty' is not loaded}%
    \renewcommand\transparent[1]{}%
  }%
  \providecommand\rotatebox[2]{#2}%
  \ifx\svgwidth\undefined%
    \setlength{\unitlength}{228.4822887bp}%
    \ifx\svgscale\undefined%
      \relax%
    \else%
      \setlength{\unitlength}{\unitlength * \real{\svgscale}}%
    \fi%
  \else%
    \setlength{\unitlength}{\svgwidth}%
  \fi%
  \global\let\svgwidth\undefined%
  \global\let\svgscale\undefined%
  \makeatother%
  \begin{picture}(1,0.74297524)%
    \put(0,0){\includegraphics[width=\unitlength]{figure7.pdf}}%
    \put(0.15206676,0.05005805){\color[rgb]{0,0,0}\makebox(0,0)[lb]{\smash{0}}}%
    \put(0.30454737,0.05005805){\color[rgb]{0,0,0}\makebox(0,0)[lb]{\smash{2}}}%
    \put(0.45702794,0.05005805){\color[rgb]{0,0,0}\makebox(0,0)[lb]{\smash{4}}}%
    \put(0.60950859,0.04856317){\color[rgb]{0,0,0}\makebox(0,0)[lb]{\smash{6}}}%
    \put(0.76273667,0.05005805){\color[rgb]{0,0,0}\makebox(0,0)[lb]{\smash{8}}}%
    \put(0.90774273,0.05005805){\color[rgb]{0,0,0}\makebox(0,0)[lb]{\smash{10}}}%
    \put(0.4066102,0.00755666){\color[rgb]{0,0,0}\makebox(0,0)[lb]{\smash{AFD angle, $\Theta$ ($^\circ$)}}}%
    \put(0.04026035,0.07889532){\color[rgb]{0,0,0}\makebox(0,0)[lb]{\smash{-2}}}%
    \put(0.04175526,0.23899988){\color[rgb]{0,0,0}\makebox(0,0)[lb]{\smash{-1}}}%
    \put(0.04922979,0.39910443){\color[rgb]{0,0,0}\makebox(0,0)[lb]{\smash{0}}}%
    \put(0.0507247,0.55920912){\color[rgb]{0,0,0}\makebox(0,0)[lb]{\smash{1}}}%
    \put(0.04922979,0.71931367){\color[rgb]{0,0,0}\makebox(0,0)[lb]{\smash{2}}}%
    \put(0.13084735,0.42240778){\color[rgb]{0,0,0}\rotatebox{90}{\makebox(0,0)[lb]{\smash{Stable}}}}%
    \put(0.1308389,0.25696283){\color[rgb]{0,0,0}\rotatebox{90}{\makebox(0,0)[lb]{\smash{Unstable}}}}%
    \put(0.02513187,0.19676749){\color[rgb]{0,0,0}\rotatebox{90}{\makebox(0,0)[lb]{\smash{Longitudinal IFC (eV/\AA$^2$)}}}}%
    \put(0.2299454,0.55669807){\color[rgb]{0,0,0}\makebox(0,0)[lb]{\smash{Ti-O in plane}}}%
    \put(0.2299454,0.51907653){\color[rgb]{0,0,0}\makebox(0,0)[lb]{\smash{Ti-O out-of plane}}}%
    \put(0.22883099,0.63194128){\color[rgb]{0,0,0}\makebox(0,0)[lb]{\smash{Sr-O out-of plane compressed}}}%
    \put(0.22883099,0.66956295){\color[rgb]{0,0,0}\makebox(0,0)[lb]{\smash{Sr-O in plane}}}%
    \put(0.22883099,0.59431974){\color[rgb]{0,0,0}\makebox(0,0)[lb]{\smash{Sr-O out-of plane stretched}}}%
  \end{picture}%
\endgroup%

%% file: figure8.pdf_tex
\begingroup%
  \makeatletter%
  \providecommand\color[2][]{%
    \errmessage{(Inkscape) Color is used for the text in Inkscape, but the package 'color.sty' is not loaded}%
    \renewcommand\color[2][]{}%
  }%
  \providecommand\transparent[1]{%
    \errmessage{(Inkscape) Transparency is used (non-zero) for the text in Inkscape, but the package 'transparent.sty' is not loaded}%
    \renewcommand\transparent[1]{}%
  }%
  \providecommand\rotatebox[2]{#2}%
  \ifx\svgwidth\undefined%
    \setlength{\unitlength}{234.58243561bp}%
    \ifx\svgscale\undefined%
      \relax%
    \else%
      \setlength{\unitlength}{\unitlength * \real{\svgscale}}%
    \fi%
  \else%
    \setlength{\unitlength}{\svgwidth}%
  \fi%
  \global\let\svgwidth\undefined%
  \global\let\svgscale\undefined%
  \makeatother%
  \begin{picture}(1,0.386427)%
    \put(0,0){\includegraphics[width=\unitlength]{figure8.pdf}}%
    \put(0.19209834,0.04831453){\color[rgb]{0,0,0}\makebox(0,0)[lb]{\smash{0}}}%
    \put(0.33780704,0.04831453){\color[rgb]{0,0,0}\makebox(0,0)[lb]{\smash{2}}}%
    \put(0.48351571,0.04831453){\color[rgb]{0,0,0}\makebox(0,0)[lb]{\smash{4}}}%
    \put(0.62922445,0.04688611){\color[rgb]{0,0,0}\makebox(0,0)[lb]{\smash{6}}}%
    \put(0.77564739,0.04831453){\color[rgb]{0,0,0}\makebox(0,0)[lb]{\smash{8}}}%
    \put(0.91421356,0.04831453){\color[rgb]{0,0,0}\makebox(0,0)[lb]{\smash{10}}}%
    \put(0.40350121,0.00736015){\color[rgb]{0,0,0}\makebox(0,0)[lb]{\smash{AFD angle, $\Theta$ (deg)}}}%
    \put(0.03893494,0.12100358){\color[rgb]{0,0,0}\makebox(0,0)[lb]{\smash{-0.04}}}%
    \put(0.03916807,0.17286627){\color[rgb]{0,0,0}\makebox(0,0)[lb]{\smash{-0.02}}}%
    \put(0.0929871,0.22472871){\color[rgb]{0,0,0}\makebox(0,0)[lb]{\smash{0}}}%
    \put(0.05055798,0.27659101){\color[rgb]{0,0,0}\makebox(0,0)[lb]{\smash{0.02}}}%
    \put(0.05032485,0.32845344){\color[rgb]{0,0,0}\makebox(0,0)[lb]{\smash{0.04}}}%
    \put(0.02447834,0.1877008){\color[rgb]{0,0,0}\rotatebox{90}{\makebox(0,0)[lb]{\smash{$\Delta$q (e)}}}}%
    \put(0.25157893,0.31671396){\color[rgb]{0,0,0}\makebox(0,0)[lb]{\smash{Sr}}}%
    \put(0.25026756,0.27865689){\color[rgb]{0,0,0}\makebox(0,0)[lb]{\smash{Ti}}}%
  \end{picture}%
\endgroup%

%% file: figure9.pdf_tex

\begingroup
  \makeatletter
  \providecommand\color[2][]{%
    \errmessage{(Inkscape) Color is used for the text in Inkscape, but the package 'color.sty' is not loaded}
    \renewcommand\color[2][]{}%
  }
  \providecommand\transparent[1]{%
    \errmessage{(Inkscape) Transparency is used (non-zero) for the text in Inkscape, but the package 'transparent.sty' is not loaded}
    \renewcommand\transparent[1]{}%
  }
  \providecommand\rotatebox[2]{#2}
  \ifx\svgwidth\undefined
    \setlength{\unitlength}{234.63591003pt}
  \else
    \setlength{\unitlength}{\svgwidth}
  \fi
  \global\let\svgwidth\undefined
  \makeatother
  \begin{picture}(1,0.89662348)%
    \put(0,0){\includegraphics[width=\unitlength]{figure9.pdf}}%
    \put(0.15909795,0.49895497){\color[rgb]{0,0,0}\makebox(0,0)[lb]{\smash{0}}}%
    \put(0.28773087,0.49855542){\color[rgb]{0,0,0}\makebox(0,0)[lb]{\smash{2}}}%
    \put(0.41636378,0.49855542){\color[rgb]{0,0,0}\makebox(0,0)[lb]{\smash{4}}}%
    \put(0.54499669,0.49895497){\color[rgb]{0,0,0}\makebox(0,0)[lb]{\smash{6}}}%
    \put(0.67453042,0.49895497){\color[rgb]{0,0,0}\makebox(0,0)[lb]{\smash{8}}}%
    \put(0.79415539,0.49895497){\color[rgb]{0,0,0}\makebox(0,0)[lb]{\smash{10}}}%
    \put(0.32626619,0.4639247){\color[rgb]{0,0,0}\makebox(0,0)[lb]{\smash{AFD angle, $\Theta$ (deg)}}}%
    \put(0.04220531,0.52531989){\color[rgb]{0,0,0}\makebox(0,0)[lb]{\smash{1.7}}}%
    \put(0.04265481,0.5949723){\color[rgb]{0,0,0}\makebox(0,0)[lb]{\smash{1.8}}}%
    \put(0.04215537,0.66462497){\color[rgb]{0,0,0}\makebox(0,0)[lb]{\smash{1.9}}}%
    \put(0.04190564,0.73427738){\color[rgb]{0,0,0}\makebox(0,0)[lb]{\smash{2.0}}}%
    \put(0.04485236,0.80393005){\color[rgb]{0,0,0}\makebox(0,0)[lb]{\smash{2.1}}}%
    \put(0.04210542,0.87358246){\color[rgb]{0,0,0}\makebox(0,0)[lb]{\smash{2.2}}}%
    \put(0.02367365,0.60564272){\color[rgb]{0,0,0}\rotatebox{90}{\makebox(0,0)[lb]{\smash{Band gap (eV)}}}}%
    \put(0.12630575,0.04917182){\color[rgb]{0,0,0}\makebox(0,0)[lb]{\smash{0}}}%
    \put(0.19025503,0.04877227){\color[rgb]{0,0,0}\makebox(0,0)[lb]{\smash{1}}}%
    \put(0.25240275,0.04877227){\color[rgb]{0,0,0}\makebox(0,0)[lb]{\smash{2}}}%
    \put(0.31635206,0.04917182){\color[rgb]{0,0,0}\makebox(0,0)[lb]{\smash{3}}}%
    \put(0.37849977,0.04877227){\color[rgb]{0,0,0}\makebox(0,0)[lb]{\smash{4}}}%
    \put(0.44154824,0.04917182){\color[rgb]{0,0,0}\makebox(0,0)[lb]{\smash{5}}}%
    \put(0.19140655,0.00729188){\color[rgb]{0,0,0}\makebox(0,0)[lb]{\smash{Ti 3d U (eV)}}}%
    \put(0.06151027,0.07432767){\color[rgb]{0,0,0}\makebox(0,0)[lb]{\smash{-4}}}%
    \put(0.06174334,0.15171938){\color[rgb]{0,0,0}\makebox(0,0)[lb]{\smash{-2}}}%
    \put(0.07293089,0.2291111){\color[rgb]{0,0,0}\makebox(0,0)[lb]{\smash{0}}}%
    \put(0.07313067,0.30650281){\color[rgb]{0,0,0}\makebox(0,0)[lb]{\smash{2}}}%
    \put(0.07289759,0.38389452){\color[rgb]{0,0,0}\makebox(0,0)[lb]{\smash{4}}}%
    \put(0.02367365,0.1179705){\color[rgb]{0,0,0}\rotatebox{90}{\makebox(0,0)[lb]{\smash{Frequency (THz)}}}}%
    \put(0.62830941,0.36242285){\color[rgb]{0,0,0}\makebox(0,0)[lb]{\smash{FE mode}}}%
    \put(0.62859243,0.32295923){\color[rgb]{0,0,0}\makebox(0,0)[lb]{\smash{AFD mode}}}%
    \put(0.51972931,0.04917182){\color[rgb]{0,0,0}\makebox(0,0)[lb]{\smash{0}}}%
    \put(0.58369378,0.04877227){\color[rgb]{0,0,0}\makebox(0,0)[lb]{\smash{1}}}%
    \put(0.64585648,0.04877227){\color[rgb]{0,0,0}\makebox(0,0)[lb]{\smash{2}}}%
    \put(0.70982088,0.04917182){\color[rgb]{0,0,0}\makebox(0,0)[lb]{\smash{3}}}%
    \put(0.77198371,0.04877227){\color[rgb]{0,0,0}\makebox(0,0)[lb]{\smash{4}}}%
    \put(0.83504717,0.04917182){\color[rgb]{0,0,0}\makebox(0,0)[lb]{\smash{5}}}%
    \put(0.58486773,0.00729188){\color[rgb]{0,0,0}\makebox(0,0)[lb]{\smash{Sr 4d U (eV)}}}%
    \put(0.88466114,0.11313968){\color[rgb]{0,0,0}\makebox(0,0)[lb]{\smash{-100}}}%
    \put(0.88466114,0.17022457){\color[rgb]{0,0,0}\makebox(0,0)[lb]{\smash{-50}}}%
    \put(0.88481098,0.2291111){\color[rgb]{0,0,0}\makebox(0,0)[lb]{\smash{0}}}%
    \put(0.88441142,0.28619599){\color[rgb]{0,0,0}\makebox(0,0)[lb]{\smash{50}}}%
    \put(0.88204739,0.34508252){\color[rgb]{0,0,0}\makebox(0,0)[lb]{\smash{100}}}%
    \put(0.88204739,0.4021674){\color[rgb]{0,0,0}\makebox(0,0)[lb]{\smash{150}}}%
    \put(0.99264152,0.11115152){\color[rgb]{0,0,0}\rotatebox{90}{\makebox(0,0)[lb]{\smash{Frequency (cm$^{-1}$)}}}}%
    \put(0.83510381,0.55155033){\color[rgb]{0,0,0}\makebox(0,0)[lb]{\smash{a)}}}%
    \put(0.44369273,0.09739392){\color[rgb]{0,0,0}\makebox(0,0)[lb]{\smash{b)}}}%
    \put(0.83734474,0.09739392){\color[rgb]{0,0,0}\makebox(0,0)[lb]{\smash{c)}}}%
  \end{picture}%
\endgroup

%% file: STO_Oct23.bbl
\begin{thebibliography}{35}%
\makeatletter
\providecommand \@ifxundefined [1]{%
 \@ifx{#1\undefined}
}%
\providecommand \@ifnum [1]{%
 \ifnum #1\expandafter \@firstoftwo
 \else \expandafter \@secondoftwo
 \fi
}%
\providecommand \@ifx [1]{%
 \ifx #1\expandafter \@firstoftwo
 \else \expandafter \@secondoftwo
 \fi
}%
\providecommand \natexlab [1]{#1}%
\providecommand \enquote  [1]{``#1''}%
\providecommand \bibnamefont  [1]{#1}%
\providecommand \bibfnamefont [1]{#1}%
\providecommand \citenamefont [1]{#1}%
\providecommand \href@noop [0]{\@secondoftwo}%
\providecommand \href [0]{\begingroup \@sanitize@url \@href}%
\providecommand \@href[1]{\@@startlink{#1}\@@href}%
\providecommand \@@href[1]{\endgroup#1\@@endlink}%
\providecommand \@sanitize@url [0]{\catcode `\\12\catcode `\$12\catcode
  `\&12\catcode `\#12\catcode `\^12\catcode `\_12\catcode `\%12\relax}%
\providecommand \@@startlink[1]{}%
\providecommand \@@endlink[0]{}%
\providecommand \url  [0]{\begingroup\@sanitize@url \@url }%
\providecommand \@url [1]{\endgroup\@href {#1}{\urlprefix }}%
\providecommand \urlprefix  [0]{URL }%
\providecommand \Eprint [0]{\href }%
\providecommand \doibase [0]{http://dx.doi.org/}%
\providecommand \selectlanguage [0]{\@gobble}%
\providecommand \bibinfo  [0]{\@secondoftwo}%
\providecommand \bibfield  [0]{\@secondoftwo}%
\providecommand \translation [1]{[#1]}%
\providecommand \BibitemOpen [0]{}%
\providecommand \bibitemStop [0]{}%
\providecommand \bibitemNoStop [0]{.\EOS\space}%
\providecommand \EOS [0]{\spacefactor3000\relax}%
\providecommand \BibitemShut  [1]{\csname bibitem#1\endcsname}%
\let\auto@bib@innerbib\@empty
\bibitem [{\citenamefont {Rabe}\ \emph {et~al.}(2007)\citenamefont {Rabe},
  \citenamefont {Triscone},\ and\ \citenamefont {Ahn}}]{Rabe:2007up}%
  \BibitemOpen
  \bibfield  {author} {\bibinfo {author} {\bibfnamefont {K.~M.}\ \bibnamefont
  {Rabe}}, \bibinfo {author} {\bibfnamefont {J.-M.}\ \bibnamefont {Triscone}},
  \ and\ \bibinfo {author} {\bibfnamefont {C.~H.}\ \bibnamefont {Ahn}},\
  }\href@noop {} {\emph {\bibinfo {title} {{Physics of Ferroelectrics: A modern
  perspective}}}}\ (\bibinfo  {publisher} {Springer},\ \bibinfo {year}
  {2007})\BibitemShut {NoStop}%
\bibitem [{\citenamefont {Bednorz}\ and\ \citenamefont
  {M{\"u}ller}(1988)}]{Bednorz:1988bw}%
  \BibitemOpen
  \bibfield  {author} {\bibinfo {author} {\bibfnamefont {J.~G.}\ \bibnamefont
  {Bednorz}}\ and\ \bibinfo {author} {\bibfnamefont {K.~A.}\ \bibnamefont
  {M{\"u}ller}},\ }\href@noop {} {\bibfield  {journal} {\bibinfo  {journal}
  {Rev Mod Phys}\ }\textbf {\bibinfo {volume} {60}},\ \bibinfo {pages} {585}
  (\bibinfo {year} {1988})}\BibitemShut {NoStop}%
\bibitem [{\citenamefont {Ramirez}(1997)}]{Ramirez:1997uh}%
  \BibitemOpen
  \bibfield  {author} {\bibinfo {author} {\bibfnamefont {A.}~\bibnamefont
  {Ramirez}},\ }\href@noop {} {\bibfield  {journal} {\bibinfo  {journal} {J.
  Phys.: Condens. Matter}\ }\textbf {\bibinfo {volume} {9}},\ \bibinfo {pages}
  {8171} (\bibinfo {year} {1997})}\BibitemShut {NoStop}%
\bibitem [{\citenamefont {Woodward}(1997{\natexlab{a}})}]{Woodward:1997ci}%
  \BibitemOpen
  \bibfield  {author} {\bibinfo {author} {\bibfnamefont {P.~M.}\ \bibnamefont
  {Woodward}},\ }\href@noop {} {\bibfield  {journal} {\bibinfo  {journal} {Acta
  Cryst. B}\ }\textbf {\bibinfo {volume} {53}},\ \bibinfo {pages} {32}
  (\bibinfo {year} {1997}{\natexlab{a}})}\BibitemShut {NoStop}%
\bibitem [{\citenamefont {Woodward}(1997{\natexlab{b}})}]{Woodward:1997js}%
  \BibitemOpen
  \bibfield  {author} {\bibinfo {author} {\bibfnamefont {P.~M.}\ \bibnamefont
  {Woodward}},\ }\href@noop {} {\bibfield  {journal} {\bibinfo  {journal} {Acta
  Cryst. B}\ }\textbf {\bibinfo {volume} {53}},\ \bibinfo {pages} {44}
  (\bibinfo {year} {1997}{\natexlab{b}})}\BibitemShut {NoStop}%
\bibitem [{\citenamefont {Eklund}\ \emph {et~al.}(2009)\citenamefont {Eklund},
  \citenamefont {Fennie},\ and\ \citenamefont {Rabe}}]{Eklund:2009hp}%
  \BibitemOpen
  \bibfield  {author} {\bibinfo {author} {\bibfnamefont {C.~J.}\ \bibnamefont
  {Eklund}}, \bibinfo {author} {\bibfnamefont {C.}~\bibnamefont {Fennie}}, \
  and\ \bibinfo {author} {\bibfnamefont {K.~M.}\ \bibnamefont {Rabe}},\
  }\href@noop {} {\bibfield  {journal} {\bibinfo  {journal} {Phys. Rev. B}\
  }\textbf {\bibinfo {volume} {79}},\ \bibinfo {pages} {220101} (\bibinfo
  {year} {2009})}\BibitemShut {NoStop}%
\bibitem [{\citenamefont {Bhattacharjee}\ \emph {et~al.}(2009)\citenamefont
  {Bhattacharjee}, \citenamefont {Bousquet},\ and\ \citenamefont
  {Ghosez}}]{Bhattacharjee:2009eb}%
  \BibitemOpen
  \bibfield  {author} {\bibinfo {author} {\bibfnamefont {S.}~\bibnamefont
  {Bhattacharjee}}, \bibinfo {author} {\bibfnamefont {E.}~\bibnamefont
  {Bousquet}}, \ and\ \bibinfo {author} {\bibfnamefont {P.}~\bibnamefont
  {Ghosez}},\ }\href@noop {} {\bibfield  {journal} {\bibinfo  {journal} {Phys.
  Rev. Lett.}\ }\textbf {\bibinfo {volume} {102}},\ \bibinfo {pages} {117602}
  (\bibinfo {year} {2009})}\BibitemShut {NoStop}%
\bibitem [{\citenamefont {Hatt}\ and\ \citenamefont
  {Spaldin}(2010)}]{Hatt:2010fi}%
  \BibitemOpen
  \bibfield  {author} {\bibinfo {author} {\bibfnamefont {A.~J.}\ \bibnamefont
  {Hatt}}\ and\ \bibinfo {author} {\bibfnamefont {N.~A.}\ \bibnamefont
  {Spaldin}},\ }\href@noop {} {\bibfield  {journal} {\bibinfo  {journal} {Phys.
  Rev. B}\ }\textbf {\bibinfo {volume} {82}},\ \bibinfo {pages} {195402}
  (\bibinfo {year} {2010})}\BibitemShut {NoStop}%
\bibitem [{\citenamefont {Rondinelli}\ and\ \citenamefont
  {Spaldin}(2009)}]{Rondinelli:2009fk}%
  \BibitemOpen
  \bibfield  {author} {\bibinfo {author} {\bibfnamefont {J.}~\bibnamefont
  {Rondinelli}}\ and\ \bibinfo {author} {\bibfnamefont {N.~A.}\ \bibnamefont
  {Spaldin}},\ }\href@noop {} {\bibfield  {journal} {\bibinfo  {journal} {Phys.
  Rev. B}\ }\textbf {\bibinfo {volume} {79}},\ \bibinfo {pages} {054409}
  (\bibinfo {year} {2009})}\BibitemShut {NoStop}%
\bibitem [{\citenamefont {Benedek}\ and\ \citenamefont
  {Fennie}(2013)}]{Benedek:2013jl}%
  \BibitemOpen
  \bibfield  {author} {\bibinfo {author} {\bibfnamefont {N.~A.}\ \bibnamefont
  {Benedek}}\ and\ \bibinfo {author} {\bibfnamefont {C.~J.}\ \bibnamefont
  {Fennie}},\ }\href@noop {} {\bibfield  {journal} {\bibinfo  {journal} {J.
  Phys. Chem. C}\ }\textbf {\bibinfo {volume} {117}},\ \bibinfo {pages} {13339}
  (\bibinfo {year} {2013})}\BibitemShut {NoStop}%
\bibitem [{\citenamefont {Shirane}\ and\ \citenamefont
  {Yamada}(1969)}]{Shirane:1969ug}%
  \BibitemOpen
  \bibfield  {author} {\bibinfo {author} {\bibfnamefont {G.}~\bibnamefont
  {Shirane}}\ and\ \bibinfo {author} {\bibfnamefont {Y.}~\bibnamefont
  {Yamada}},\ }\href@noop {} {\bibfield  {journal} {\bibinfo  {journal} {Phys
  Rev}\ }\textbf {\bibinfo {volume} {177}},\ \bibinfo {pages} {858} (\bibinfo
  {year} {1969})}\BibitemShut {NoStop}%
\bibitem [{\citenamefont {Cowley}\ \emph {et~al.}(1969)\citenamefont {Cowley},
  \citenamefont {Buyers},\ and\ \citenamefont {Dolling}}]{Cowley:1969ul}%
  \BibitemOpen
  \bibfield  {author} {\bibinfo {author} {\bibfnamefont {R.~A.}\ \bibnamefont
  {Cowley}}, \bibinfo {author} {\bibfnamefont {W.}~\bibnamefont {Buyers}}, \
  and\ \bibinfo {author} {\bibfnamefont {G.}~\bibnamefont {Dolling}},\
  }\href@noop {} {\bibfield  {journal} {\bibinfo  {journal} {Solid State
  Communications}\ }\textbf {\bibinfo {volume} {7}},\ \bibinfo {pages} {181}
  (\bibinfo {year} {1969})}\BibitemShut {NoStop}%
\bibitem [{\citenamefont {Glazer}(1972)}]{Glazer:1972eb}%
  \BibitemOpen
  \bibfield  {author} {\bibinfo {author} {\bibfnamefont {A.~M.}\ \bibnamefont
  {Glazer}},\ }\href@noop {} {\bibfield  {journal} {\bibinfo  {journal} {Acta
  Cryst. B}\ }\textbf {\bibinfo {volume} {28}},\ \bibinfo {pages} {3384}
  (\bibinfo {year} {1972})}\BibitemShut {NoStop}%
\bibitem [{\citenamefont {Zhong}\ and\ \citenamefont
  {Vanderbilt}(1996)}]{Zhong:1996vq}%
  \BibitemOpen
  \bibfield  {author} {\bibinfo {author} {\bibfnamefont {W.}~\bibnamefont
  {Zhong}}\ and\ \bibinfo {author} {\bibfnamefont {D.}~\bibnamefont
  {Vanderbilt}},\ }\href@noop {} {\bibfield  {journal} {\bibinfo  {journal}
  {Phys. Rev. B}\ }\textbf {\bibinfo {volume} {53}},\ \bibinfo {pages} {5047}
  (\bibinfo {year} {1996})}\BibitemShut {NoStop}%
\bibitem [{\citenamefont {Vanderbilt}\ and\ \citenamefont
  {Zhong}(1998)}]{Vanderbilt:1998fs}%
  \BibitemOpen
  \bibfield  {author} {\bibinfo {author} {\bibfnamefont {D.}~\bibnamefont
  {Vanderbilt}}\ and\ \bibinfo {author} {\bibfnamefont {W.}~\bibnamefont
  {Zhong}},\ }\href@noop {} {\bibfield  {journal} {\bibinfo  {journal}
  {Ferroelectrics}\ }\textbf {\bibinfo {volume} {206}},\ \bibinfo {pages} {181}
  (\bibinfo {year} {1998})}\BibitemShut {NoStop}%
\bibitem [{\citenamefont {M{\"u}ller}\ and\ \citenamefont
  {Burkard}(1979)}]{Muller:1979wa}%
  \BibitemOpen
  \bibfield  {author} {\bibinfo {author} {\bibfnamefont {K.~A.}\ \bibnamefont
  {M{\"u}ller}}\ and\ \bibinfo {author} {\bibfnamefont {H.}~\bibnamefont
  {Burkard}},\ }\href@noop {} {\bibfield  {journal} {\bibinfo  {journal} {Phys.
  Rev. B}\ }\textbf {\bibinfo {volume} {19}},\ \bibinfo {pages} {3593}
  (\bibinfo {year} {1979})}\BibitemShut {NoStop}%
\bibitem [{\citenamefont {Kresse}\ and\ \citenamefont
  {Hafner}(1993)}]{Kresse:1993ty}%
  \BibitemOpen
  \bibfield  {author} {\bibinfo {author} {\bibfnamefont {G.}~\bibnamefont
  {Kresse}}\ and\ \bibinfo {author} {\bibfnamefont {J.}~\bibnamefont
  {Hafner}},\ }\href@noop {} {\bibfield  {journal} {\bibinfo  {journal} {Phys.
  Rev. B}\ }\textbf {\bibinfo {volume} {47}},\ \bibinfo {pages} {558} (\bibinfo
  {year} {1993})}\BibitemShut {NoStop}%
\bibitem [{\citenamefont {Kresse}\ and\ \citenamefont
  {Hafner}(1994)}]{Kresse:1994us}%
  \BibitemOpen
  \bibfield  {author} {\bibinfo {author} {\bibfnamefont {G.}~\bibnamefont
  {Kresse}}\ and\ \bibinfo {author} {\bibfnamefont {J.}~\bibnamefont
  {Hafner}},\ }\href@noop {} {\bibfield  {journal} {\bibinfo  {journal} {Phys.
  Rev. B}\ }\textbf {\bibinfo {volume} {49}},\ \bibinfo {pages} {14251}
  (\bibinfo {year} {1994})}\BibitemShut {NoStop}%
\bibitem [{\citenamefont {Kresse}\ and\ \citenamefont
  {Furthmuller}(1996{\natexlab{a}})}]{Kresse:1996vk}%
  \BibitemOpen
  \bibfield  {author} {\bibinfo {author} {\bibfnamefont {G.}~\bibnamefont
  {Kresse}}\ and\ \bibinfo {author} {\bibfnamefont {J.}~\bibnamefont
  {Furthmuller}},\ }\href@noop {} {\bibfield  {journal} {\bibinfo  {journal}
  {Comp. Mater. Sci.}\ }\textbf {\bibinfo {volume} {6}},\ \bibinfo {pages} {15}
  (\bibinfo {year} {1996}{\natexlab{a}})}\BibitemShut {NoStop}%
\bibitem [{\citenamefont {Kresse}\ and\ \citenamefont
  {Furthmuller}(1996{\natexlab{b}})}]{Kresse:1996vf}%
  \BibitemOpen
  \bibfield  {author} {\bibinfo {author} {\bibfnamefont {G.}~\bibnamefont
  {Kresse}}\ and\ \bibinfo {author} {\bibfnamefont {J.}~\bibnamefont
  {Furthmuller}},\ }\href@noop {} {\bibfield  {journal} {\bibinfo  {journal}
  {Phys. Rev. B}\ }\textbf {\bibinfo {volume} {54}},\ \bibinfo {pages} {11169}
  (\bibinfo {year} {1996}{\natexlab{b}})}\BibitemShut {NoStop}%
\bibitem [{\citenamefont {Bl{\"o}chl}(1994)}]{Blochl:1994uk}%
  \BibitemOpen
  \bibfield  {author} {\bibinfo {author} {\bibfnamefont {P.~E.}\ \bibnamefont
  {Bl{\"o}chl}},\ }\href@noop {} {\bibfield  {journal} {\bibinfo  {journal}
  {Phys. Rev. B}\ }\textbf {\bibinfo {volume} {50}},\ \bibinfo {pages} {17953}
  (\bibinfo {year} {1994})}\BibitemShut {NoStop}%
\bibitem [{\citenamefont {Kresse}\ and\ \citenamefont
  {Joubert}(1999)}]{Kresse:1999wc}%
  \BibitemOpen
  \bibfield  {author} {\bibinfo {author} {\bibfnamefont {G.}~\bibnamefont
  {Kresse}}\ and\ \bibinfo {author} {\bibfnamefont {D.}~\bibnamefont
  {Joubert}},\ }\href@noop {} {\bibfield  {journal} {\bibinfo  {journal} {Phys.
  Rev. B}\ }\textbf {\bibinfo {volume} {59}},\ \bibinfo {pages} {1758}
  (\bibinfo {year} {1999})}\BibitemShut {NoStop}%
\bibitem [{\citenamefont {Perdew}\ \emph {et~al.}(1996)\citenamefont {Perdew},
  \citenamefont {Burke},\ and\ \citenamefont {Ernzerhof}}]{Perdew:1996iq}%
  \BibitemOpen
  \bibfield  {author} {\bibinfo {author} {\bibfnamefont {J.~P.}\ \bibnamefont
  {Perdew}}, \bibinfo {author} {\bibfnamefont {K.}~\bibnamefont {Burke}}, \
  and\ \bibinfo {author} {\bibfnamefont {M.}~\bibnamefont {Ernzerhof}},\
  }\href@noop {} {\bibfield  {journal} {\bibinfo  {journal} {Phys. Rev. Lett.}\
  }\textbf {\bibinfo {volume} {77}},\ \bibinfo {pages} {3865} (\bibinfo {year}
  {1996})}\BibitemShut {NoStop}%
\bibitem [{\citenamefont {Perdew}\ \emph {et~al.}(2008)\citenamefont {Perdew},
  \citenamefont {Ruzsinszky}, \citenamefont {Csonka}, \citenamefont {Vydrov},
  \citenamefont {Scuseria}, \citenamefont {Constantin}, \citenamefont {Zhou},\
  and\ \citenamefont {Burke}}]{Perdew:2008fa}%
  \BibitemOpen
  \bibfield  {author} {\bibinfo {author} {\bibfnamefont {J.~P.}\ \bibnamefont
  {Perdew}}, \bibinfo {author} {\bibfnamefont {A.}~\bibnamefont {Ruzsinszky}},
  \bibinfo {author} {\bibfnamefont {G.}~\bibnamefont {Csonka}}, \bibinfo
  {author} {\bibfnamefont {O.}~\bibnamefont {Vydrov}}, \bibinfo {author}
  {\bibfnamefont {G.}~\bibnamefont {Scuseria}}, \bibinfo {author}
  {\bibfnamefont {L.}~\bibnamefont {Constantin}}, \bibinfo {author}
  {\bibfnamefont {X.}~\bibnamefont {Zhou}}, \ and\ \bibinfo {author}
  {\bibfnamefont {K.}~\bibnamefont {Burke}},\ }\href@noop {} {\bibfield
  {journal} {\bibinfo  {journal} {Phys. Rev. Lett.}\ }\textbf {\bibinfo
  {volume} {100}},\ \bibinfo {pages} {136406} (\bibinfo {year}
  {2008})}\BibitemShut {NoStop}%
\bibitem [{\citenamefont {Monkhorst}\ and\ \citenamefont
  {Pack}(1976)}]{Monkhorst:1976ta}%
  \BibitemOpen
  \bibfield  {author} {\bibinfo {author} {\bibfnamefont {H.~J.}\ \bibnamefont
  {Monkhorst}}\ and\ \bibinfo {author} {\bibfnamefont {J.~D.}\ \bibnamefont
  {Pack}},\ }\href@noop {} {\bibfield  {journal} {\bibinfo  {journal} {Phys.
  Rev. B}\ }\textbf {\bibinfo {volume} {13}},\ \bibinfo {pages} {5188}
  (\bibinfo {year} {1976})}\BibitemShut {NoStop}%
\bibitem [{\citenamefont {Togo}\ \emph {et~al.}(2008)\citenamefont {Togo},
  \citenamefont {Oba},\ and\ \citenamefont {Tanaka}}]{Togo:2008jt}%
  \BibitemOpen
  \bibfield  {author} {\bibinfo {author} {\bibfnamefont {A.}~\bibnamefont
  {Togo}}, \bibinfo {author} {\bibfnamefont {F.}~\bibnamefont {Oba}}, \ and\
  \bibinfo {author} {\bibfnamefont {I.}~\bibnamefont {Tanaka}},\ }\href@noop {}
  {\bibfield  {journal} {\bibinfo  {journal} {Phys. Rev. B}\ }\textbf {\bibinfo
  {volume} {78}},\ \bibinfo {pages} {134106} (\bibinfo {year}
  {2008})}\BibitemShut {NoStop}%
\bibitem [{\citenamefont {Wahl}\ \emph {et~al.}(2008)\citenamefont {Wahl},
  \citenamefont {Vogtenhuber},\ and\ \citenamefont {Kresse}}]{Wahl:2008ea}%
  \BibitemOpen
  \bibfield  {author} {\bibinfo {author} {\bibfnamefont {R.}~\bibnamefont
  {Wahl}}, \bibinfo {author} {\bibfnamefont {D.}~\bibnamefont {Vogtenhuber}}, \
  and\ \bibinfo {author} {\bibfnamefont {G.}~\bibnamefont {Kresse}},\
  }\href@noop {} {\bibfield  {journal} {\bibinfo  {journal} {Phys. Rev. B}\
  }\textbf {\bibinfo {volume} {78}},\ \bibinfo {pages} {104116} (\bibinfo
  {year} {2008})}\BibitemShut {NoStop}%
\bibitem [{\citenamefont {Evarestov}\ \emph {et~al.}(2011)\citenamefont
  {Evarestov}, \citenamefont {Blokhin}, \citenamefont {Gryaznov}, \citenamefont
  {Kotomin},\ and\ \citenamefont {Maier}}]{Evarestov:2011iv}%
  \BibitemOpen
  \bibfield  {author} {\bibinfo {author} {\bibfnamefont {R.~A.}\ \bibnamefont
  {Evarestov}}, \bibinfo {author} {\bibfnamefont {E.}~\bibnamefont {Blokhin}},
  \bibinfo {author} {\bibfnamefont {D.}~\bibnamefont {Gryaznov}}, \bibinfo
  {author} {\bibfnamefont {E.~A.}\ \bibnamefont {Kotomin}}, \ and\ \bibinfo
  {author} {\bibfnamefont {J.}~\bibnamefont {Maier}},\ }\href@noop {}
  {\bibfield  {journal} {\bibinfo  {journal} {Phys. Rev. B}\ }\textbf {\bibinfo
  {volume} {83}},\ \bibinfo {pages} {134108} (\bibinfo {year}
  {2011})}\BibitemShut {NoStop}%
\bibitem [{\citenamefont {El-Mellouhi}\ \emph {et~al.}(2013)\citenamefont
  {El-Mellouhi}, \citenamefont {Brothers}, \citenamefont {Lucero},
  \citenamefont {Bulik},\ and\ \citenamefont {Scuseria}}]{ElMellouhi:2013bu}%
  \BibitemOpen
  \bibfield  {author} {\bibinfo {author} {\bibfnamefont {F.}~\bibnamefont
  {El-Mellouhi}}, \bibinfo {author} {\bibfnamefont {E.}~\bibnamefont
  {Brothers}}, \bibinfo {author} {\bibfnamefont {M.}~\bibnamefont {Lucero}},
  \bibinfo {author} {\bibfnamefont {I.}~\bibnamefont {Bulik}}, \ and\ \bibinfo
  {author} {\bibfnamefont {G.}~\bibnamefont {Scuseria}},\ }\href@noop {}
  {\bibfield  {journal} {\bibinfo  {journal} {Phys. Rev. B}\ }\textbf {\bibinfo
  {volume} {87}},\ \bibinfo {pages} {035107} (\bibinfo {year}
  {2013})}\BibitemShut {NoStop}%
\bibitem [{\citenamefont {Haeni}\ \emph {et~al.}(2004)\citenamefont {Haeni},
  \citenamefont {Irvin}, \citenamefont {Chang}, \citenamefont {Uecker},
  \citenamefont {Reiche}, \citenamefont {Li}, \citenamefont {Choudhury},
  \citenamefont {Tian}, \citenamefont {Hawley}, \citenamefont {Craigo},
  \citenamefont {Tagantsev}, \citenamefont {Pan}, \citenamefont {Streiffer},
  \citenamefont {Chen}, \citenamefont {Kirchoefer}, \citenamefont {Levy},\ and\
  \citenamefont {Schlom}}]{Haeni:2004gj}%
  \BibitemOpen
  \bibfield  {author} {\bibinfo {author} {\bibfnamefont {J.~H.}\ \bibnamefont
  {Haeni}}, \bibinfo {author} {\bibfnamefont {P.}~\bibnamefont {Irvin}},
  \bibinfo {author} {\bibfnamefont {W.}~\bibnamefont {Chang}}, \bibinfo
  {author} {\bibfnamefont {R.}~\bibnamefont {Uecker}}, \bibinfo {author}
  {\bibfnamefont {P.}~\bibnamefont {Reiche}}, \bibinfo {author} {\bibfnamefont
  {Y.~L.}\ \bibnamefont {Li}}, \bibinfo {author} {\bibfnamefont
  {S.}~\bibnamefont {Choudhury}}, \bibinfo {author} {\bibfnamefont
  {W.}~\bibnamefont {Tian}}, \bibinfo {author} {\bibfnamefont {M.~E.}\
  \bibnamefont {Hawley}}, \bibinfo {author} {\bibfnamefont {B.}~\bibnamefont
  {Craigo}}, \bibinfo {author} {\bibfnamefont {A.~K.}\ \bibnamefont
  {Tagantsev}}, \bibinfo {author} {\bibfnamefont {X.~Q.}\ \bibnamefont {Pan}},
  \bibinfo {author} {\bibfnamefont {S.~K.}\ \bibnamefont {Streiffer}}, \bibinfo
  {author} {\bibfnamefont {L.~Q.}\ \bibnamefont {Chen}}, \bibinfo {author}
  {\bibfnamefont {S.~W.}\ \bibnamefont {Kirchoefer}}, \bibinfo {author}
  {\bibfnamefont {J.}~\bibnamefont {Levy}}, \ and\ \bibinfo {author}
  {\bibfnamefont {D.~G.}\ \bibnamefont {Schlom}},\ }\href@noop {} {\bibfield
  {journal} {\bibinfo  {journal} {Nature}\ }\textbf {\bibinfo {volume} {430}},\
  \bibinfo {pages} {758} (\bibinfo {year} {2004})}\BibitemShut {NoStop}%
\bibitem [{\citenamefont {Choi}\ \emph {et~al.}(2004)\citenamefont {Choi},
  \citenamefont {Biegalski}, \citenamefont {Li}, \citenamefont {Sharan},
  \citenamefont {Schubert}, \citenamefont {Uecker}, \citenamefont {Reiche},
  \citenamefont {Chen}, \citenamefont {Pan}, \citenamefont {Gopalan},
  \citenamefont {Chen}, \citenamefont {Schlom},\ and\ \citenamefont
  {Eom}}]{Choi:2004it}%
  \BibitemOpen
  \bibfield  {author} {\bibinfo {author} {\bibfnamefont {K.~J.}\ \bibnamefont
  {Choi}}, \bibinfo {author} {\bibfnamefont {M.}~\bibnamefont {Biegalski}},
  \bibinfo {author} {\bibfnamefont {Y.~L.}\ \bibnamefont {Li}}, \bibinfo
  {author} {\bibfnamefont {A.}~\bibnamefont {Sharan}}, \bibinfo {author}
  {\bibfnamefont {J.}~\bibnamefont {Schubert}}, \bibinfo {author}
  {\bibfnamefont {R.}~\bibnamefont {Uecker}}, \bibinfo {author} {\bibfnamefont
  {P.}~\bibnamefont {Reiche}}, \bibinfo {author} {\bibfnamefont {Y.~B.}\
  \bibnamefont {Chen}}, \bibinfo {author} {\bibfnamefont {X.~Q.}\ \bibnamefont
  {Pan}}, \bibinfo {author} {\bibfnamefont {V.}~\bibnamefont {Gopalan}},
  \bibinfo {author} {\bibfnamefont {L.~Q.}\ \bibnamefont {Chen}}, \bibinfo
  {author} {\bibfnamefont {D.~G.}\ \bibnamefont {Schlom}}, \ and\ \bibinfo
  {author} {\bibfnamefont {C.~B.}\ \bibnamefont {Eom}},\ }\href@noop {}
  {\bibfield  {journal} {\bibinfo  {journal} {Science}\ }\textbf {\bibinfo
  {volume} {306}},\ \bibinfo {pages} {1005} (\bibinfo {year}
  {2004})}\BibitemShut {NoStop}%
\bibitem [{\citenamefont {Zhong}\ \emph {et~al.}(1994)\citenamefont {Zhong},
  \citenamefont {King-Smith},\ and\ \citenamefont {Vanderbilt}}]{Zhong:1994ui}%
  \BibitemOpen
  \bibfield  {author} {\bibinfo {author} {\bibfnamefont {W.}~\bibnamefont
  {Zhong}}, \bibinfo {author} {\bibfnamefont {R.~D.}\ \bibnamefont
  {King-Smith}}, \ and\ \bibinfo {author} {\bibfnamefont {D.}~\bibnamefont
  {Vanderbilt}},\ }\href@noop {} {\bibfield  {journal} {\bibinfo  {journal}
  {Phys. Rev. Lett.}\ }\textbf {\bibinfo {volume} {72}},\ \bibinfo {pages}
  {3618} (\bibinfo {year} {1994})}\BibitemShut {NoStop}%
\bibitem [{\citenamefont {Lasota}\ \emph {et~al.}(1997)\citenamefont {Lasota},
  \citenamefont {Wang}, \citenamefont {Yu},\ and\ \citenamefont
  {Krakauer}}]{Lasota:1997dq}%
  \BibitemOpen
  \bibfield  {author} {\bibinfo {author} {\bibfnamefont {C.}~\bibnamefont
  {Lasota}}, \bibinfo {author} {\bibfnamefont {C.-Z.}\ \bibnamefont {Wang}},
  \bibinfo {author} {\bibfnamefont {R.}~\bibnamefont {Yu}}, \ and\ \bibinfo
  {author} {\bibfnamefont {H.}~\bibnamefont {Krakauer}},\ }\href@noop {}
  {\bibfield  {journal} {\bibinfo  {journal} {Ferroelectrics}\ }\textbf
  {\bibinfo {volume} {194}},\ \bibinfo {pages} {109} (\bibinfo {year}
  {1997})}\BibitemShut {NoStop}%
\bibitem [{\citenamefont {Rondinelli}\ \emph {et~al.}(2009)\citenamefont
  {Rondinelli}, \citenamefont {Eidelson},\ and\ \citenamefont
  {Spaldin}}]{Rondinelli:2009tq}%
  \BibitemOpen
  \bibfield  {author} {\bibinfo {author} {\bibfnamefont {J.~M.}\ \bibnamefont
  {Rondinelli}}, \bibinfo {author} {\bibfnamefont {A.}~\bibnamefont
  {Eidelson}}, \ and\ \bibinfo {author} {\bibfnamefont {N.~A.}\ \bibnamefont
  {Spaldin}},\ }\href@noop {} {\bibfield  {journal} {\bibinfo  {journal} {Phys.
  Rev. B}\ }\textbf {\bibinfo {volume} {79}},\ \bibinfo {pages} {205119}
  (\bibinfo {year} {2009})}\BibitemShut {NoStop}%
\bibitem [{\citenamefont {Berger}\ \emph {et~al.}(2011)\citenamefont {Berger},
  \citenamefont {Fennie},\ and\ \citenamefont {Neaton}}]{Berger:2011jv}%
  \BibitemOpen
  \bibfield  {author} {\bibinfo {author} {\bibfnamefont {R.~F.}\ \bibnamefont
  {Berger}}, \bibinfo {author} {\bibfnamefont {C.~J.}\ \bibnamefont {Fennie}},
  \ and\ \bibinfo {author} {\bibfnamefont {J.~B.}\ \bibnamefont {Neaton}},\
  }\href@noop {} {\bibfield  {journal} {\bibinfo  {journal} {Phys. Rev. Lett.}\
  }\textbf {\bibinfo {volume} {107}},\ \bibinfo {pages} {146804} (\bibinfo
  {year} {2011})}\BibitemShut {NoStop}%
\end{thebibliography}%
